\documentclass[usenatbib]{mnras}

\usepackage{amssymb}

\usepackage{newtxtext,newtxmath}
\usepackage{appendix}

\usepackage[T1]{fontenc}

\DeclareRobustCommand{\VAN}[3]{#2}
\let\VANthebibliography\thebibliography
\def\thebibliography{\DeclareRobustCommand{\VAN}[3]{##3}\VANthebibliography}

\usepackage{graphicx}	
\usepackage{amsmath}	
\usepackage{color}
\usepackage{threeparttable, tablefootnote}  
\usepackage{stfloats}
\usepackage{tabularx}

\newcommand{\eg}{{\sl e.g.},}   
        
\newcommand{\rtwoh}{R_{\rm 200c}}        
\newcommand{\mtwoh}{M_{\rm 200c}}

\newcommand{\mstarBCG}{M_{\star,\rm BCG}}

\newcommand{\kms}{{\rm \, km~s}\ensuremath{^{-1}}}
\newcommand{\hinv}{\ensuremath{\, h^{-1}}}%
\newcommand{\msol}{\ensuremath{\, {\rm M}_\odot}}

\newcommand{\mpc}{\ensuremath{\, {\rm Mpc}}}

\newcommand{\Nsat}{\ensuremath{N_{\rm sat}}}

\newcommand{\mlim}{M_{\rm lim}}        

\newcommand{\lcdm}{$\Lambda$CDM}
\newcommand{\sigmaeight}{\ensuremath{\sigma_8}}
\newcommand{\omegam}{\ensuremath{\Omega_{\rm m}}}

\newcommand{\Seight}{\ensuremath{S_8}}

\definecolor{bleudefrance}{rgb}{0.19, 0.55, 0.91}
\definecolor{purple}{RGB}{128, 0, 128}
\newcommand*{\gus}[1]{\textcolor{bleudefrance}{\textsf{Gus: #1}}}

\newcommand{\Smeas}{\ensuremath{S_{\rm obs}}}
\newcommand{\Sobs}{\ensuremath{S}}
\newcommand{\sobs}{\ensuremath{s}}
\newcommand{\sobsbar}{\ensuremath{\overline{\sobs}}}

\newcommand{\sigmamu}{\ensuremath{\sigma_{\mu}}}
\newcommand{\campi}{\ensuremath{\varpi}}

\newcommand{\errmubar}{\ensuremath{\epsilon_{\langle M \rangle}}}
\newcommand{\errVarmu}{\ensuremath{\epsilon_{{\rm Var}\mu}}}

\title{Cluster Cosmology Redux: A Compact Representation for the Halo Mass Function }
\author{Cameron E. Norton, Fred C. Adams and August E. Evrard}
\date{\today}

\begin{document}

\maketitle

\begin{abstract}
Massive halos hosting groups and clusters of galaxies imprint coherent, arcminute-scale features across the  spectrophotometric sky, especially optical-IR clusters of galaxies, distortions in the sub-mm CMB, and extended sources of X-ray emission.  Statistical modeling of such features often rely upon the evolving space-time density of dark matter halos -- the halo mass function (HMF) -- as a common theoretical ground for cosmological, astrophysical and fundamental physics studies. 
We propose a compact (eight parameter) representation of the HMF with readily interpretable parameters that stem from polynomial expansions, first in terms of log-mass, then expanding those coefficients similarly in redshift.  We demonstrate good ($\sim \! 5\%$) agreement of this form, referred to as the dual-quadratic (DQ-HMF), with Mira-Titan N-body emulator estimates for halo masses above $10^{13.7} \hinv\msol$ over the redshift range $0.1 < z < 1.5$, present best-fit parameters for a Planck 2018 cosmology, and present parameter variation in the $\sigmaeight - \omegam$ plane. 
Convolving with a minimal mass--observable relation (MOR) yields closed-form expressions for counts, mean mass, and mass variance of cluster samples characterized by some observable property.  
Performing information-matrix forecasts of potential parameter constraints from existing and future surveys under different levels of systematic uncertainties, 
we demonstrate the potential for percent-level constraints on model parameters by an LSST-like optical cluster survey of 300,000 clusters and a richness--mass variance of $0.3^2$. Even better constraints could potentially be achieved by a survey with one-tenth the sample size but with a reduced selection property variance of $0.1^2$.  Potential benefits and extensions to the basic MOR parameterization are discussed. 

\end{abstract}


\section{Introduction}\label{sec:Intro}

The evolving population of galaxy clusters on the sky is a cosmological diagnostic whose value has been 
recognized since the era when 4-m class telescopes opened the study of clusters at redshifts above 0.5 \citep{Gunn1986, Peebles1989, Evrard1989}.  The massive halos that host groups and clusters of galaxies represent a rare event tail of hierarchical structure formation that is sensitive to both the growth rate of linear structure \citep{White1993sigma8} and the nature of the initial fluctuation power spectrum \citep{Dalal2008nonGaussianity}.

Constraints on cosmological parameters forecast for deep and wide cluster samples two decades ago \citep[{\rm e.g.,}][]{Haiman2001, Holder2001, BattyeWeller2003} are now emerging from cluster selection methods based on features observed in optical-IR surveys  \citep{Gladders2007RCSclustercosmo, Rozo2010SDSSclustercosmo, Rykoff2014redMaPPerI,  Gonzales2019WISEclusters, Abdullah2020SDSSclustercosmo, Costanzi2020DESY1, Miyatake2021HSCcosmo, Aguena2021WAZP, Wen2022DES+unWISE, Maturi2023JPASamicoClusters}, thermal Sunyaev-Zel'dovich (SZ) effect on the cosmic microwave background \citep{Sehgal2011ACTclustercosmo, deHaan2016SPTclustercosmo, PlanckXXIV2016clustercosmo, Bocquet2019SPTclustercosmo} and extended X-ray emission \citep{Bohringer2007REXCESS,  Vikhlinin2009, Mehrtens2012XCS, Pierre2016XXLintro, Pacaud2018XXLcosmo, Chitham2020CODEXcosmo, Chiu2023eFEDScosmo}.  Local cluster counts are sensitive to the current linear power spectrum amplitude, $\sigmaeight$, and matter density parameter, $\omegam$, particularly through the combination, $\Seight \equiv \sigmaeight (\omegam/0.3)^{0.5}$ \citep{AllenEvrardMantz2011ARAA}. 

Cosmological constraints from the aforementioned studies are sometimes inconsistent.  Dark Energy Survey Year One (DES-Y1) analysis, based on counts and mean lensing masses in four richness and three redshift bins with a total sample size of 6500 clusters, find $\Seight =  0.65 \pm 0.04$, significantly ($4\sigma$) below the $0.830 \pm 0.013$ value from Planck 2018 CMB analysis \citep{Planck2018Cosmology}.  In contrast KIDS-DR3 cluster population analysis \citep{Lesci2022KIDSclusterCosmo}, based on a data vector similar to that of DES-Y1 derived from an optical sample of nearly 3700 clusters, yields $\Seight = 0.78 \pm 0.04$, $1\sigma$ consistent with the Planck CMB value. 

Within a given cosmology, formulating expectations for cluster counts and aggregate lensing masses of samples selected on some observable property is challenged by several sources of systematic uncertainty.  The physical extent of massive halos and their preference to form in large-scale overdense regions of the cosmic web creates source confusion; the virial regions of $M > 3 \times 10^{13} \msol$ halos hosting groups and clusters of galaxies cover one-third of the \lcdm\ sky within $z<1.5$ \citep{VoitEvrardBryan2001}.  Projection tends to boost intrinsic properties \citep[\eg][]{White2002SZsims, Cohn2007MillenRSclusters, Costanzi2019opticalProjection}, but the fact that the effects of projected structure on optical, X-ray, and SZ measurements will generally differ reinforces the value of multi-wavelength cluster sample analysis.  


The statistical relationship between the bulk observable properties of a halo, such as its X-ray temperature, gas or stellar mass, or galaxy richness\footnote{Background-subtracted count, often of red galaxies, within a characteristic radius, \citep[\eg][]{Rozo2009lambda}.}, and its true total mass is another source of uncertainty \citep[\eg][]{Salvati2020, Wu2021}.  This relationship connects the sky+redshift-space abundance of a property-selected cluster population to the space-time density of massive halos.
The differential form of the latter point density, known as halo mass function (HMF), 
is now well characterized in the space of standard \lcdm\ cosmological parameters by large N-body simulation campaigns \citep[\eg][and references therein]{MiraTitan2021}.  

A convolution of the HMF with the mass--observable relation (MOR) is the basis of survey statistical expectations, and a power-law mean with log-normal variance is a canonical MOR form motivated by cosmological hydrodynamics simulations \citep[\eg][and references therein]{BryanNorman1998scaling,  Angulo2012MillXXXscaling, Farahi2018BMscaling, Anbajagane2020stellarProps}.  Over a wide dynamic range in mass, a single power law mean may be insufficient, especially for hot gas properties \citep{Farahi2018BMscaling, Pop2022TNGscaling}, and the variance may also be mass-dependent \citep{Anbajagane2020stellarProps}.  Extensions to accommodate such behavior are discussed in \S\ref{sec:discExtensions}.  

The convolution naturally joins cluster astrophysics, encapsulated by the MOR, to the (primarily) cosmology-driven HMF, and the parameter couplings of these spaces have been explored previously \citep[][hereafter, E14]{Evrard2014}. The model we present here extends previous work by letting the HMF shape parameters vary continuously with redshift.  Essentially, E14 introduced approximate HMF forms at a fixed epoch in order to develop expressions for conditional statistics of samples selected by an observable property. This present work develops a continuous space-time representation of the differential space density at high halo masses with the goal of constructing a compact, interpretable form of the HMF.

In this paper, we first show that a simple, eight parameter representation captures the near-field ($z<1.5$) group/cluster HMF derived by the Mira-Titan universe ensemble \citep{MiraTitan2021}.  Coupled with a log-normally distributed MOR, we derive closed-form expressions for both the evolving space density and the log-mean selected mass of the group/cluster population as a function of the selection property and redshift.  

With those ingredients, we then perform an information matrix (IM)\footnote{We omit using the proper name associated with this method due to \href{https://nautil.us/how-eugenics-shaped-statistics-238014}{that person's embrace of eugenics principles}.} analysis to explore the ability of current and future cluster surveys --- specifically, those based on counts and mean gravitational lensing mass as a function of a chosen selection property and redshift --- to constrain the parameters of this HMF form.  The forecasts require information on the expected uncertainty in lensing mass measurements as well as the uncertainty in the MOR variance, and we consider both current estimates and future advances in our projections.  

Why consider cluster cosmology as an HMF-centric exercise?  The first reason is that the evolving shape of the HMF is interesting in its own right, as it contains information on both $w$CDM parameters and other cosmological physics, including light neutrino masses \citep{Marulli2011Neutrinos, Costanzi2013NeutrinoHMF, Hagstotz2019ModGrav, Hernandez-AguayoMilleniumTNG, Euclid2022NeutrinoSimComp}, modified gravity models \citep{Schmidt2009ModGrav, Zhao2011ModGrav, Cataneo2016ModGrav, Arnold2019ModGrav, Hagstotz2019ModGrav, Mitchell2021ModGrav}, and non-Guassian initial fluctuations \citep{Matarrese2000NonGauss, Sefusatti2007nonGauss, Grossi2009nonGauss, Pillepich2010nonGauss, LoVerde2011nonGauss, HarrisonColes2011nonGauss,  Jung2023QuijotePNG, Coulton2023QuijotePNG}.  The astrophysics of galaxy formation affects the  HMF shape in non-trivial ways that continue to be studied by cosmological hydrodynamics simulations \citep{Stanek2009baryonHMF, Cui2012baryonHMF, Cui2014baryonHMF, Martizzi2014baryonHMF, Cusworth2014baryonHMF, Castro2021baryonHMF, Schaye2023Flamingo}.

Another reason is that HMF-centric analyses can exploit increasingly tight constraints on the differential comoving volume element, $dV/dz$, from baryon acoustic oscillations \citep[BAO,][]{eBOSS2021, DES2022BAOdistance} and Type Ia supernovae \citep[SN,][]{Guy2010SNdistance, DES2019Y3SNcosmo, Brout2022PantheonSNdistance, Mitra2023LSST_SNdistance}.  In current methods of cluster survey analysis, $dV/dz$ is left free to vary in the space of \lcdm\ parameters.

From a practical perspective, a compact representation for the HMF at galaxy cluster scales can serve as a consistency check among cluster samples selected at different wavelengths.  As the common ground that underlies all cluster samples, the inferred HMF needs to be consistent across surveys and independent of the chosen sample selection property.  
An important feature of our model is that it naturally incorporates multiple, intrinsically correlated physical properties \citep{Mulroy2019LoCuSS, Farahi2019LoCuSS}. 

We are far from the first to emphasize the HMF shape. The original study of \citet{BahcallCen1993} used counts of nearby clusters selected by optical and X-ray properties to directly estimate the HMF of the low-redshift universe.  That work benefited from the insensitivity of nearby volume to cosmological mean density parameters.  Subsequent studies derived HMF estimates from X-ray samples \citep{Reiprich2002, Bohringer2017} or optical cluster samples using galaxy richness \citep{Bahcall2003} or velocity dispersion \citep{Pisani2003, Rines2007, Rines2008} as a proxy for mass.  The statistical power of these samples was limited by their moderate sample sizes, typically several hundred systems. 


Compact HMF representations already exist, but historically they have been expressed in terms of a similarity variable, $\sigma^2(M)$, the rms amplitude of linear perturbations smoothed on a Lagrangian scale $R \propto M^{1/3}$ \citep{PressSchechter1974}.  An assumption about cosmology is required to convert these forms to a function of mass. 
The Sheth-Tormen (ST) form \citep{ShethMoTormen2001} is a popular example,  and constraints on the parameters of this model have been published from analysis of magnified images of sub-mm galaxies \citep{Cueli2022} and from counts of GAMA groups and clusters \citep{DriverHMF2022}.  

Because the ST model represents a non-linear function of the similarity variable rather than mass directly, its parameters are difficult to interpret.  By expressing the HMF directly in terms of halo mass and redshift, the eight free parameters of our model (see Table~\ref{tab:HMFparams} below) have clear interpretations as coefficients of polynomial expansions.  

The aims of this paper are twofold.  We first demonstrate the model's ability to reproduce LCDM sky counts from the Mira-Titan emulator\footnote{In fact, \citet{MiraTitan2021} use a piecewise quadratic in log-mass as the basis of their emulator method.} in the space of fluctuation amplitude, $\sigma_8$, and matter density parameter, $\omegam$ \citep{MiraTitan2021}, and provide Planck 2018 model parameters.  We then apply an information matrix approach to estimate potential constraints on DQ-HMF parameters from idealized cluster surveys patterned after existing \citep[DES-Y1,][]{Costanzi2020DESY1} and future \citep[LSST,][]{Chisari2019ApJS_LSSTcode} galaxy cluster surveys.  
The parameter forecasts employ cluster counts and mean weak lensing masses, each derived within finite ranges of selection property and redshift, along with an additional input on the degree of scatter in log-mass at fixed value of the selection property.  

In \S\ref{sec:Methods} we detail the model's structure, demonstrate its utility at capturing emulator predictions in the space of $\{\sigma_8,\omegam\}$.  Expressions for counts and mean mass as a function of an observable property are presented in \S\ref{sec:OBSsample}, and the IM elements we employ for survey analysis are also defined there.  Forecasts of parameter constraints from existing and planned cluster surveys are presented in \S\ref{sec:Results}.  In \S\ref{sec:Discussion} we discuss ideas for implementing the model and review how massive halos tie to many non-Gaussian LSS signatures. Benefits of selecting a sub-sample with reduced property variance are made explicit in \S\ref{sec:discSelectionML}. An appendix offers a three-parameter toy model that helps illustrate the key role of MOR variance. 
 
We employ a cosmology with matter density ${\Omega_m = 0.311}$, baryon density ${\Omega_b = 0.0489}$, Hubble constant ${H_0 = 67.7 \kms \mpc^{-1}}$, primordial spectral index ${n_s = 0.967}$, and power spectrum normalization ${\sigma_8 = 0.810}$, values derived by Planck2018 CMB+BAO analysis. Our measure of halo mass is $\mtwoh$, the mass defined by a mean interior spherical overdensity of 200 times the critical density, $\rho_c(z)$, and we express this mass in units of $\hinv \msol$, where $h= H_0/100 \kms \,\mpc^{-1}$.  Our spatial density unit for the HMF is $h^3 \mpc^{-3}$.  
The IM analysis is patterned after optical survey samples but is generalizable to samples selected by other properties.  Relative to X-ray and SZ selection, optical samples have the benefit of distance estimation from spectroscopic or photometric redshifts. We ignore distance uncertainties, as the redshift bins we employ are much wider than typical uncertainties \citep{Rykoff2014redMaPPerI, Rykoff2016DES-SVredmapper, Maturi2023JPASamicoClusters}.


\section{Methods}\label{sec:Methods}

A key component of \lcdm\ structure formation is an initially Gaussian random density field whose amplitude grows due to gravity.  A spherical collapse model \citep{GunnGott1972} argues for a linearly evolved perturbation amplitude threshold at which halos form.  Combining these elements, the HMF was originally derived by \citet{PressSchechter1974} as a derivative with respect to scale of the fraction of mass in the universe that satisfies the collapse condition.  At the highest masses, for which only extreme peaks in the density field can have collapsed, this fraction is an error function with large argument, and its derivative leads to a steeply falling HMF with mass.  At lower masses, where more modest-sized perturbations can collapse, the HMF transitions to nearly a power-law form.  

The model presented in \S\ref{sec:HMFform} represents the high mass portion by the tail of a Gaussian in log-mass, meaning the log of the HMF scales as a negative quadratic with mass.  These three coefficients are themselves expanded as polynomials with redshift. While E14 included a cubic log-HMF representation, we defer that approach to future work as the quadratic form captures much of the information available in cluster counts, as we show in \S~\ref{sec:MiraTitanFits} below.  HMF parameter values in the space of $\{ \sigma_8, \omegam \}$ are presented in \S\ref{sec:HMFvaryCosmo}.

\subsection{A compact form for the cluster-scale HMF } \label{sec:HMFform}

The HMF describes the comoving spatial number density of halos as a function of mass and redshift.  Considering a small volume, $dV$, at some redshift, $z$, the probability that the center of a halo of mass, $M$, lies within that volume defines the differential HMF 
\begin{equation} \label{eq:hmfdefn}
dp \equiv \left[ \frac{dn (M,z)}{d \! \ln \! M} \right] d \! \ln \! M \, dV.  
\end{equation}
The convention of number density per logarithmic unit of mass used above implies that the HMF has dimension of inverse volume per logarithmic unit of mass.  We express the HMF amplitude in units of $h^3 \mpc^{-3}$.

We introduce an eight parameter model that employs low-order polynomial forms in log-mass and redshift. 
Letting $\mu \equiv \ln(M/M_p)$, where $M_p$ is a characteristic (pivot) mass scale, we begin with the E14 quadratic form for the log of the HMF, 
\begin{equation} \label{eq:hmfsimp}
    \ln \left[ \frac{dn(\mu, z)}{d\mu} \right] = -\sum_{i=0}^2 \  \frac{1}{i!} \ \beta_i(z) \ \mu^i .
\end{equation}
The characteristic mass is essentially the pivot scale of a quadratic expansion of the log HMF, with $\beta_0(z)$ is the normalization, $\beta_1(z)$ the local slope, and $\beta_2(z)$ the curvature of $\ln [dn/d \mu]$.  We choose a pivot mass of $10^{14.3} \hinv \msol$ and apply this form for $M \ge 10^{13.7} \hinv \msol$.  Below this mass scale the HMF transitions to a pure power-law form \citep[\eg][and references therein]{ShethMoTormen2001}.  

The  explicit negative sign on the RHS of equation~\eqref{eq:hmfsimp} is used so that the $\beta_i(z)$ parameters take on positive values.  We choose this form over an explicit Gaussian representation because the latter would imply a global representation over a very wide mass range.  Instead, we are operating on a relatively narrow mass range, roughy 1.5 decades wide, out on the Gaussian's tail, so a description using canonical location and width of a normal distribution is not as useful or meaningful.

\begin{table}
    \centering
\begin{tabularx}{\columnwidth}{c l l}
    {\bf Parameter } &  {\bf Definition}  & {\bf Value(s) } \\
     \hline \\[-6 truept]
    $\beta_{i}(z) $ & HMF evolving shape in $\mu$, $i \! \in \! [0,2] $ & see Fig.~\ref{fig:betafit} \\
    $\beta_{i,n}$ & normalization of $\beta_i$ at $z_p$ & \{12.32, 2.26, 0.75\}  \\
    $\beta_{i,z}$ & redshift gradient of $\beta_i$ at $z_p$ &  \{2.38, 1.35, 0.53\} \\
    $\beta_{i,z2}$ & redshift curvature of $\beta_i$ at $z_p$ &  \{1.39, 0.45\} \\ [2 truept]
    $M_p$ & pivot mass & $10^{14.3} \hinv \msol$ \\
    $z_p$ & pivot redshift & 0.5 \\
    $\mlim$ & minimum fit mass & $10^{13.7} \hinv \msol$  \\
\end{tabularx}
    \caption{Summary of DQ-HMF model parameters. The $\beta_i(z)$ terms represent negatives of the normalization, slope and curvature of the log-space HMF at redshift $z$ using $\mu = \ln (M/M_p)$, equation~\eqref{eq:hmfsimp}, and units of $h^3 \mpc^{-3}$. Rows two through four list the eight core HMF parameters; elements of the polynomial redshift expansions of the $\beta_i(z)$ terms around $z_p$, equations \eqref{eq:betaz} and \eqref{eq:betatwoz}.  Values for these parameters are determined by fitting to \lcdm\ Mira-Titan emulator predictions listed in Table~\ref{tab:lcdmBetas}. The final three rows list our choices of pivot locations in mass and redshift as well as the minimum mass scale of the Mira-Titan fits.}
    \label{tab:HMFparams}
\end{table}

The E14 analysis used only $\beta_i$ values defined at a few specific redshifts.  We extend that work by allowing the first pair of coefficients to run as quadratic functions of $(1+z)$, 
\begin{equation} \label{eq:betaz}
    \beta_i(z)  =  \beta_{i,n} + \beta_{i,z} \,  (z-z_p) + \frac{1}{2} \beta_{i,z2} \, (z-z_p)^2 \ \ ; \  i\in\{0,1\}, 
\end{equation}
where $z_p$ is a pivot redshift which we take to be $0.5$.  The mass curvature evolves linearly with redshift, 
\begin{equation} \label{eq:betatwoz}
    \beta_2(z)  =  \beta_{2,n} + \beta_{2,z} \,  (z-z_p) .
\end{equation}
Hereafter, we refer to equations~\eqref{eq:hmfsimp}, \eqref{eq:betaz} and \eqref{eq:betatwoz} as the dual-quadratic (DQ-HMF) model.  

Table~\ref{tab:lcdmBetas} summarizes the DQ-HMF parameters for the default \lcdm\ Planck2018 cosmology.
The following section describes how we obtained these values using the Mira-Titan emulator.  

\begin{table}
	\centering
\begin{tabular}{ c | c } 
 \textbf{Parameter} &  \textbf{Value}  \\ 
 \hline
 $\beta_{0,n}$  & 12.32  \\ 
 $\beta_{0,z}$ & 2.38 \\ 
 $\beta_{0,z2}$ & 1.39  \\ [2 truept]
 $\beta_{1, n}$ & 2.26 \\
 $\beta_{1, z}$ & 1.35 \\
 $\beta_{1, z2}$ & 0.45 \\ [2 truept]
 $\beta_{2, n}$ & 0.75 \\
 $\beta_{2, z}$ & 0.53 
\end{tabular}
   \caption{DQ-HMF parameters of the Mira-Titan \lcdm\ model. The $\beta_0$ normalization at $z_p=0.5$ is equivalent to a space density of $10^{-5.85} \mpc^{-3}$ for Hubble parameter $h=0.677$. } \label{tab:lcdmBetas}
\end{table}

\subsection{Fitting to Mira-Titan expectations}\label{sec:MiraTitanFits}

\begin{figure}
    \centering
        \vspace{-12 truept}
    \includegraphics[width = \columnwidth]{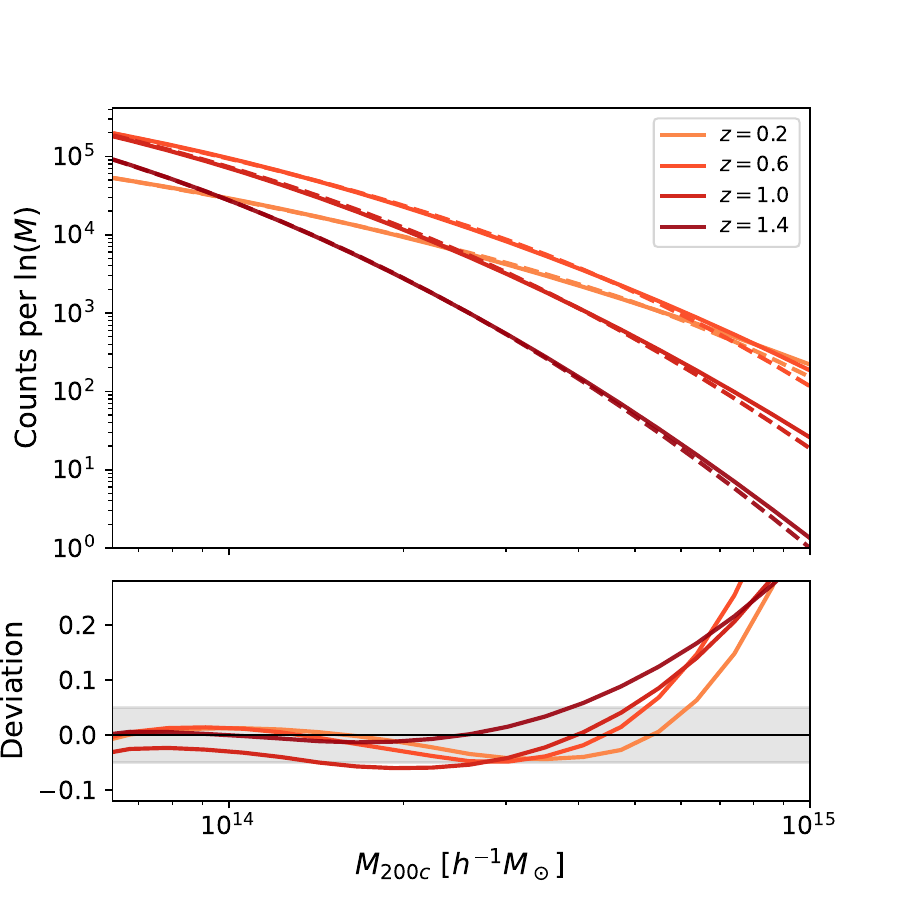}
    \vspace{-18truept}
    \caption{The upper panel shows DQ-HMF fits (solid) to the Mira-Titan emulator expectations (dashed) for counts of halos above $10^{13.7} \hinv \msol$ centered at redshifts shown in the legend over the Rubin-LSST area of $18,000$ deg$^2$.  A total of 280 sampled counts --- 20 mass bins in each of fourteen redshift shells covering the interval $0.1<z<1.5$ --- are used to fit the eight parameters of the model (see Table~\ref{tab:lcdmBetas}); we show only a subset for clarity. Poisson uncertainties applied to each binned count yield a model that best fits lower masses.  The lower panel displays the fractional deviation of the fits, $N_{\rm DQ}/N_{\rm MiraTitan} -1$, with the grey band highlighting $5\%$ agreement.  Baryon effects associated with galaxy formation drive deviations at this level or larger  \citep[\eg][]{Castro2021baryonHMF}, and the emulator itself is uncertain at the 10\%-level at $10^{15}\hinv\msol$ \citep{MiraTitan2021}.  
    }
    \label{fig:HMFplot}
\end{figure}

We evaluate the model using \lcdm\ expectations based on the Mira-Titan emulator \citep{MiraTitan2021} using a process guided by expectations for the LSST survey sky area of $18,000$ deg$^2$ \citep{2019LSSTreferenceDesign}.  We use fourteen redshift bins ranging from $z_{\rm min}=0.1$ to $z_{\rm max}=1.5$, each of width $\Delta z = 0.1$. At the central redshift, $z_j$, of each bin, we evaluate the Mira-Titan HMF and convert it to a differential number density function for an LSST-like sky area, 
\begin{equation} \label{eq:dNdmu}
    \frac{dN(\mu, z_j)}{d\mu} =  \frac{dn(\mu, z_j)}{d\mu} \ \Delta V_j, 
\end{equation}
where $\Delta V_j$ is the volume of an 18,000 sq degree survey between redshifts $z_j - 0.05$ and $z_j+0.05$.  We then integrate this form to obtain counts per bin in twenty $\mu$-bins between masses of $10^{13.7} \hinv \msol$ and $10^{15} \hinv \msol$.  We assign a Poisson uncertainty to each bin, and obtain best-fit parameters by minimizing $\chi^2$ across the combined set of 280 sampled count values.  This approach emphasizes fitting at lower masses, the range that provides the majority of information in cosmological surveys.\footnote{Indeed, \citet{Wu2021} employ counts above a single threshold, rather than differential counts, in their forecasting of $\Seight$ error from future surveys.}


Figure~\ref{fig:HMFplot} compares the eight-parameter DQ-HMF differential model counts to the Mira-Titan emulator values.  For clarity, we show four redshifts selected from the full range used in the fit; other redshifts behave similarly.  The differences between the DQ form and the emulator expectations are below 5\% at masses $< 5 \times 10^{14} \hinv \msol$, increasing to tens of percent at the highest masses.  

To contextualize the differences in the lower panel we first note that the fits are best at the lowest masses, where the information content is highest.  In addition, the finite volumes of the Mira-Titan N-body ensemble yield an HMF uncertainty of $\sim \! 10\%$ at $10^{15}\hinv\msol$ \citep{MiraTitan2021}.  Finally, the emulator is based on universes in which the clustered matter is purely collisionless matter (dark matter with an optional minority neutrino component).  The gravitational back-reaction effects of baryons cycling through the process of compact object formation can drive HMF deviations larger than $5\%$ over the mass and redshift range shown \citep{Stanek2009baryonHMF, Cui2012baryonHMF, Cui2014baryonHMF, Martizzi2014baryonHMF, Cusworth2014baryonHMF, Castro2021baryonHMF, Schaye2023Flamingo}.  Given these uncertainties, the DQ-HMF model can be considered a sufficient representation of the cluster population in the late universe.

\begin{figure}
    \centering
        \vspace{-12 truept}
    \includegraphics[width = 0.9\columnwidth]{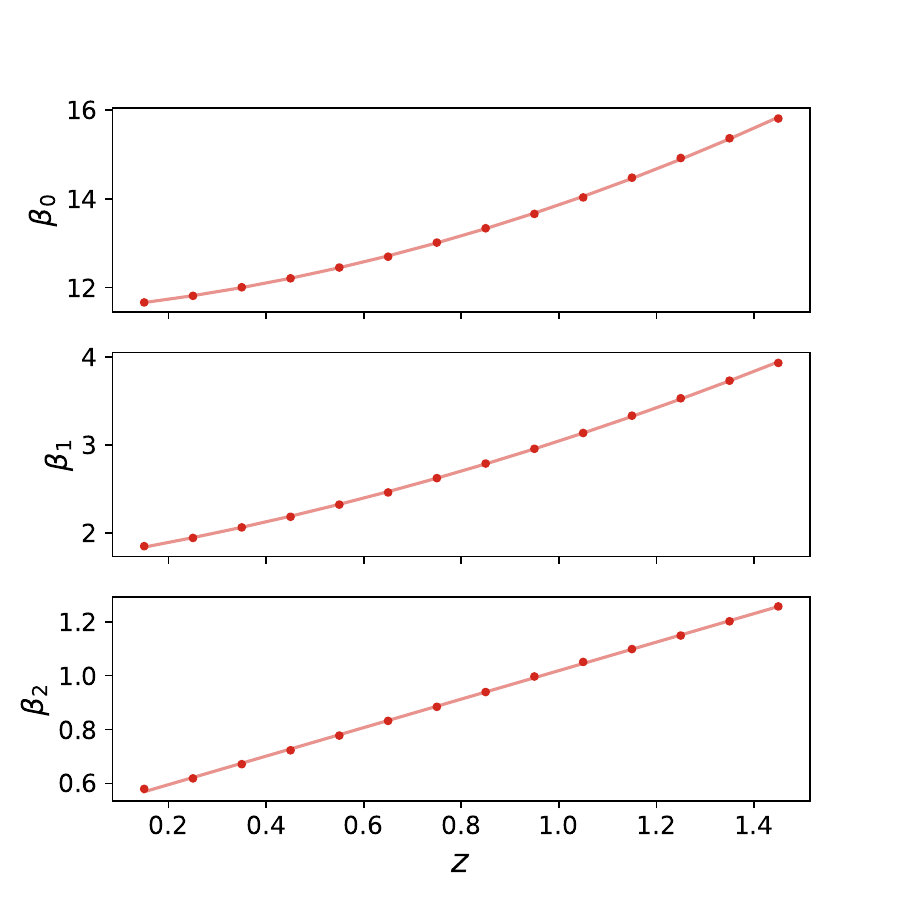}
    \vspace{-14truept}
    \caption{DQ-HMF model parameters, $\beta_i(z)$, derived from fitting Mira-Titan sky count expectations in 0.1-wide redshift shells centered at the redshifts given by the points.  A Planck 2018  \lcdm\ cosmology is assumed.  Lines show the fits to the redshift-continuous forms, equations~\eqref{eq:betaz} and \eqref{eq:betatwoz}, with parameter values given in Table~\ref{tab:lcdmBetas}.    
    }\label{fig:betafit}
\end{figure}

The model parameters resulting from the Mira-Titan emulator fits for a Planck 2018 cosmology are listed in Table~\ref{tab:lcdmBetas}. Points in Figure~\ref{fig:betafit} show $\beta_i(z)$ values determined by fitting the HMF at each sampled redshift, while lines show the redshift-continuous DQ fit with quadratic behavior of the HMF normalization and slope and linear behavior of the HMF curvature.  Halos at the pivot mass scale become increasingly rare with increasing redshift --- the normalization varies by roughly a factor of 100 over the redshift range shown --- and the HMF shape at the pivot mass becomes both steeper and more strongly curved at earlier times.  

In mapping to observable properties, the value of the slope, $\beta_1(z)$, is particularly important as it controls the amplitude of a convolution-induced bias (often referred to as Eddington bias) discussed below.  The local slope at the pivot mass of $10^{14.3} \hinv \msol$ steepens from $-2$ at $z=0.2$ to $-4$ at $z=1.5$, implying that the magnitude of this bias will grow by a factor of two over this redshift range.

\begin{figure}
    \centering
    \vspace{-12 truept}
    \includegraphics[width = 0.9\columnwidth]{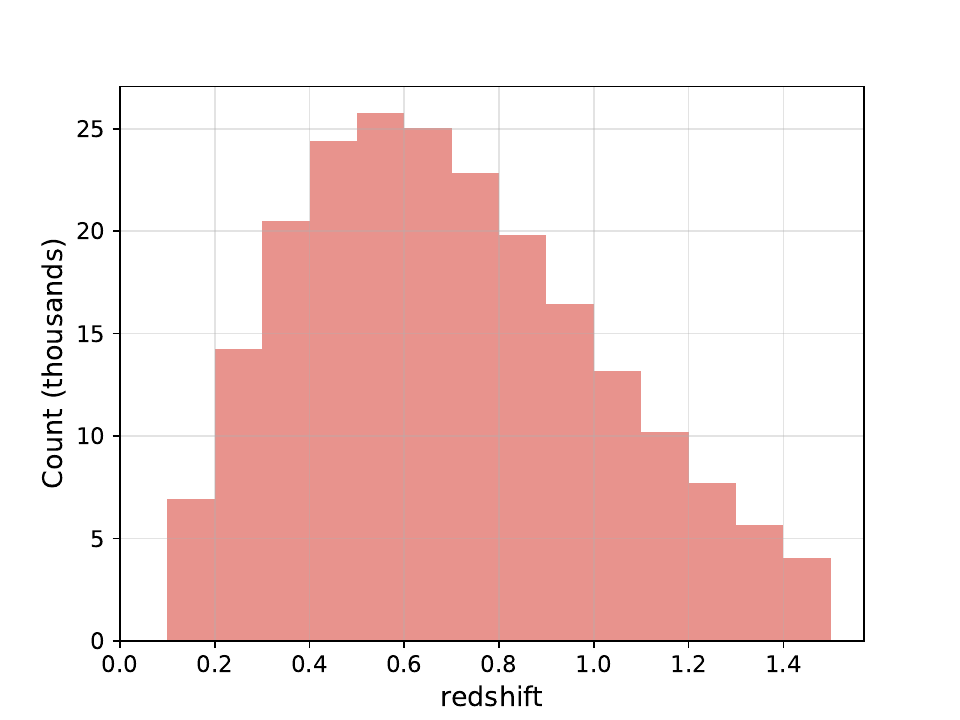}
    \vspace{-6 truept}
    \caption{Anticipated LSST-area (18000 deg$^2$) halo counts with masses, $\mtwoh \ge 10^{13.7} \hinv \msol$, in 0.1-wide redshift bins covering the range, $0.1 < z < 0.5$.  The total population of 225,000 peaks near the pivot redshift, $z_p=0.5$.    
    }
    \label{fig:zbinCounts}
\end{figure}

In the IM analysis below we explore two idealized cases, one patterned after the existing DES-Y1 cluster sample, which covers roughly 5000 deg$^2$ over redshifts, $0.2 < z < 0.65$ and another patterned after the wider, 18000 deg$^2$, and deeper LSST survey.   
Figure~\ref{fig:zbinCounts} shows the DQ model expectations for LSST survey counts of massive halos with $\mtwoh \ge 10^{13.7} \hinv \msol$ in 0.1-wide redshift bins.  A quarter million such halos should lie in the range $0.1 < z < 1.5$, with the population strongly peaked near our chosen pivot redshift of $0.5$.

\subsection{Relating HMF shape to cosmological parameters}\label{sec:HMFvaryCosmo}

The DQ-HMF shape is generic in \lcdm\ cosmologies.  
Here we use the same Mira-Titan sky expectation fitting process to map how DQ-HMF parameters vary in the canonical cluster cosmology plane of $\{ \omegam, \ \sigmaeight \}$.  
All other cosmological parameters are held constant in this exercise.  

\begin{figure*}  
    \centering
    \includegraphics[width = 2.\columnwidth]{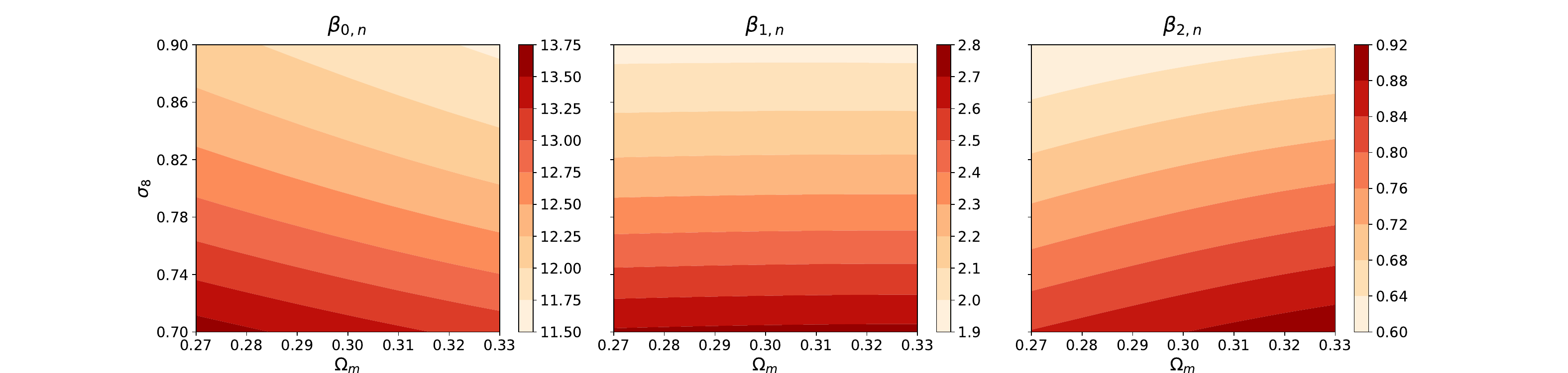}
      \includegraphics[width = 2.\columnwidth]{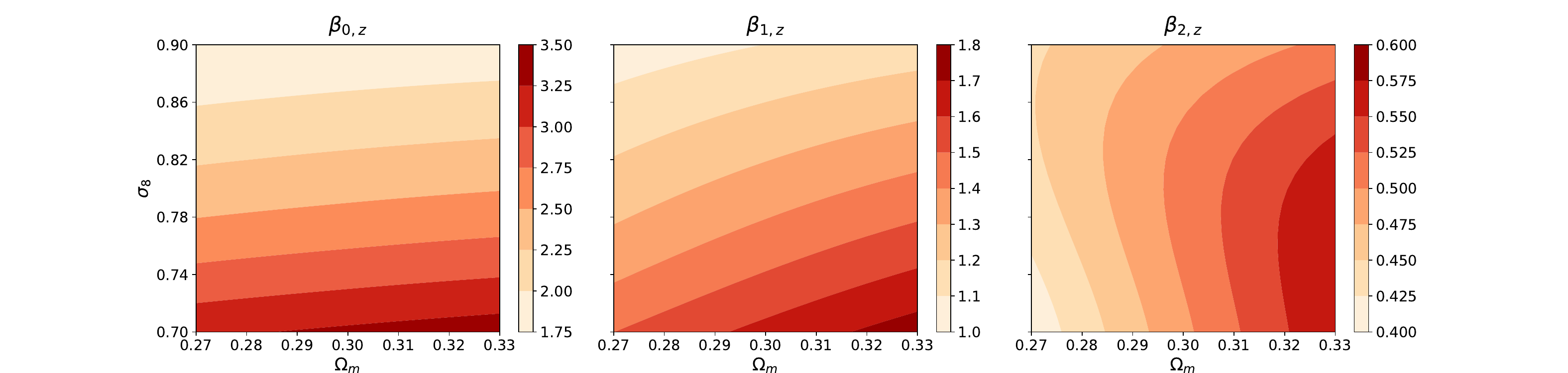}
    \includegraphics[width = 2.\columnwidth]{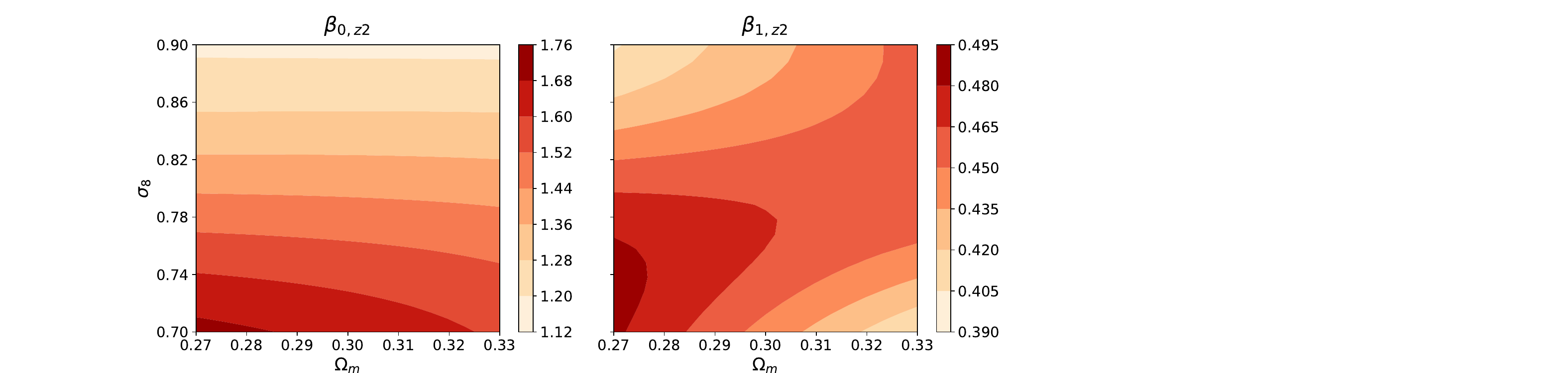}
    \caption{Contours showing DQ-HMF parameter values in the $\sigma_8$ and $\omegam$ plane.  Values are made positive by definition, equation~\eqref{eq:hmfdefn}, and all use the same pivot mass and redshift given in Table~\ref{tab:HMFparams}.  The HMF normalization, $\beta_{0,n}$ (top left panel), follows the familiar negative slope traditionally derived from cluster counts. The mass gradient, $\beta_{1,n}$ (top middle), is mainly sensitive to $\sigma_8$ while the curvature, $\beta_{2,n}$ (top right), adds information somewhat orthogonal to that of $\beta_{0,n}$.  The redshift gradient and curvature terms in the middle and lower rows display a range of behaviors, with the normalization terms sensitive only to $\sigmaeight$ and the curvature's redshift derivative is sensitive primarily to $\omegam$. The highest-order parameter, $\beta_{1,z2}$, shows non-monotonic behavior within a narrow range of values. A $20 \times 20$ sampling grid was used in the domain shown.
    }\label{fig:betacosmo}
\end{figure*}

Figure \ref{fig:betacosmo} shows the resultant behavior of the HMF shape parameters, with the top row showing normalizations, $\beta_{i,n}$, at the pivot redshift, the middle the gradients with redshift, $\beta_{i,z}$, and the bottom row the redshift curvature values for the HMF normalization and mass gradient, $\beta_{0,z2}$ and $\beta_{1,z2}$, respectively.  (Recall that the HMF curvature evolves only linearly with redshift, meaning $\beta_{2,z2} = 0$.)

Since the amplitude at the pivot redshift, $\beta_{0,n}$, is the primary controller of counts, it is not surprising that its contours tend to follow loci of $\sigma_8 \sqrt{\omegam} \simeq {\rm const.}$ in the top-left panel.  The negative of the HMF log-mass slope at $z_p$ (top middle panel) is sensitive only to $\sigma_8$, reducing from $2.8$ to $1.9$ as $\sigma_8$ increases from $0.7$ to $0.9$.  The negative of the mass curvature at the pivot redshift (upper right) behaves somewhat orthogonal to $\beta_{0,n}$, with values ranging from 0.6 to 0.9 over the range shown.  

The rate at which the HMF shape shifts over time depends is also dependent on cosmology.  The middle row of Figure~\ref{fig:betacosmo} shows redshift gradients of the mass expansion terms.  The gradient of the normalization, $\beta_{0,z}$, is highly sensitive to $\sigma_8$, scaling inversely from a low of 1.8 at $\sigma_8=0.9$ to a high of 3.5 at $\sigma_8=0.7$.  The redshift evolution of the local HMF slope, $\beta_{1,z}$, exhibits sensitivity similar to that the curvature normalization, $\beta_{2,n}$, with an amplitude variation of nearly a factor of two.  

The three terms of the highest order, $\beta_{0,z2}$, $\beta_{1,z2}$ and $\beta_{2,z}$, display mildly non-linear behaviors in the space of $\sigma_8$ and $\omegam$.  Both the redshift gradient of the HMF curvature, $\beta_{2,z}$, and the second redshift derivative of the HMF slope, $\beta_{1,z2}$, primarily depend on $\omegam$, but the latter shifts behavior at high $\sigma_8$. Not surprisingly, the highest-order parameters are anticipated to be the least well constrained in our IM analysis below.  

Note that the features displayed in Figure~\ref{fig:betacosmo} emerge from a model employing a fixed pivot mass and redshift.  Alternative choices, such as scaling the pivot mass with $\omegam$, would lead to slightly different outcomes.  Given that Planck CMB+BAO analysis limits the matter density to within a few percent ($\omegam = 0.3111 \pm  0.0065$) \citep{Planck2018Cosmology}, the shifts in the practical regions of these panels would be quite modest in size. 

\subsubsection{Massive neutrinos}\label{sec:neutrinos}


The Mira-Titan emulator allows for a non-zero neutrino mass.  Using a total neutrino mass $\sum m_\nu = 0.5$~eV while keeping all other parameters constant, we find reductions in the HMF curvature and in several redshift gradients terms of order $-0.05$.  The largest shift of $-0.1$ occurs in the rate of change of the HMF slope, $\beta_{1,z}$.  In terms of the effect on the linear growth rate of perturbations, this is roughly equivalent to reducing $\omegam$ by 0.02.  

Recent CMB lensing analysis by the ACT collaboration \citep{ACT2023cosmoLensing} limits the neutrino family total mass to 0.12~eV at $95\%$ confidence, equivalent to $\Omega_\nu h^2 = 0.001$.  For this smaller neutrino mass, DQ-HMF parameters shift at the level of 0.01 or smaller.  This level of error is potentially achievable in future surveys, albeit under optimistic conditions that would take many years to develop, as discussed in \S\ref{sec:Discussion}.  

%

\section{Observable Features of Cluster Samples}\label{sec:OBSsample}

Because the true 3D mass measure of the theoretical HMF is not directly observable, proxies that correlate with that mass measure are required.  We use a minimal MOR based on a power-law relation with log-normal scatter, a common assumption of many survey analysis models \citep{Rozo2010SDSSclustercosmo, Sehgal2011ACTclustercosmo,  deHaan2016SPTclustercosmo, PlanckXXIV2016clustercosmo, Bocquet2019SPTclustercosmo, Abdullah2020SDSSclustercosmo, Chiu2022eFeDSscaling, Lesci2022KIDSclusterCosmo}.  Generically motivated by central limit theorem arguments \citep[see, \eg][for an application to star formation]{AdamsFatuzzo1996IMFpaper}, this form is also measured in total gas and stellar mass statistics of halos realized by cosmological hydrodynamics simulations \citep{Farahi2018BMscaling, Truong2018XrayScaling,  Anbajagane2020stellarProps}.  X-ray scaling relations \citep{Pratt2009REXCESS} and lensing analysis of CAMIRA clusters \citep{Chiu2020CAMIRA_MOR} support this form empirically.  Generalizations of this approach are discussed in \S\ref{sec:Discussion}.  

The MOR model is described in \S\ref{sec:MORdefn}, followed by expressions for counts, mean mass and mass variance for samples selected by an observed property (\S\ref{sec:DataVector}).  The ingredients of our IM analysis are presented in \S\ref{sec:IMModel}. The first three rows of Table~\ref{tab:Otherparams} list the parameters needed to describe the MOR: a slope, normalization and variance.  The last three rows introduce data quality measures used in the IM analysis.

\begin{table}
    \centering
\begin{tabularx}{\columnwidth}{c l l}
    {\bf Parameter } &  {\bf Definition}  & {\bf Value(s) } \\
     \hline \\[-6 truept]
    $\campi$ & MOR normalization & 3.1 \\
    $\alpha$ & MOR slope & 1.0  \\
    $\sigma^2$ & MOR variance & $(0.3)^2$ \\
    $\errmubar$ & fractional error in mean mass & see Table~\ref{tab:QualityCases} \\  
    $\errVarmu$ & fractional error in mass variance & see Table~\ref{tab:QualityCases}   
\end{tabularx}
    \caption{Non-HMF parameters. The first three rows describe the mass--observable relation (MOR), equations~\eqref{eq:MORPDF} and \eqref{eq:MORmean}.  The variance, $\sigma^2$, is in the observed property variance at fixed mass; its inverse, the mass variance at fixed observed property, $\sigmamu^2$, is given by equation~\eqref{eq:muVar}.  The two bottom rows are assumed fractional errors in mean mass and mass variance used in the IM analysis.
    }
    \label{tab:Otherparams}
\end{table}

\subsection{Mass-conditioned property likelihood}\label{sec:MORdefn}

Let $\Sobs$ be the observable property used for sample selection, made dimensionless by the choice of a convenient reference unit, and let $\sobs \equiv \ln(\Sobs)$.  The MOR kernel is assumed to be Normal, 
\begin{equation}\label{eq:MORPDF}
    P(\sobs|\mu) = \mathcal{N}(\sobsbar(\mu), \sigma^2) =  \frac{1}{\sqrt{2\pi}\sigma}\exp\left\{- \, \frac{ [ \sobs-\sobsbar(\mu)] ^2}{2\sigma^2}\right\}, 
\end{equation}
where $\sigma^2$ is the variance in $\sobs$ at fixed halo mass.  

The mean selection property scales as a power-law in mass, meaning linearly in log-space,
\begin{equation}\label{eq:MORmean}
    \sobsbar(\mu)= \campi + \alpha \, \mu.  
\end{equation}
While carrying value as a mass proxy, $\sobs$ is not a perfect indicator of mass.  We consider only cases where $\alpha \ne 0$ and $\sigma^2 > 0$.  At fixed variance, steeper proxies are better at selecting mass (see equation~\eqref{eq:muVar} below).  In most practical cases of bulk observable properties, such as galaxy count or velocity dispersion or X-ray gas mass, the simple maxim that ``bigger is bigger'' holds, and so we generally expect that $\alpha > 0$.\footnote{There may be potential exceptions to $\alpha > 0$ scaling, such as the total mass of cold phase gas within the halo.}

For survey forecasting purposes, we assign values to the MOR parameters given in the right column of Table~\ref{tab:Otherparams}.  The normalization, $\campi = 3.1$, is equivalent to an optical richness, $\lambda = 22$, at the pivot mass scale of $10^{14.3} \hinv\msol$, and we assume a slope of unity.  Both values are consistent with the mass--richness relation of HSC clusters \citep{Murata2019HSCmassrichness}.  Other studies have found somewhat different values \citep[see][and references therein]{Abdullah2022massrichness} but we do not seek to resolve those differences here.  The variance of $0.3^2$ is consistent with estimates derived from X-ray observations of DES-Y1 clusters \citep{Farahi2019DES}.  

Although, in general, $\campi$, $\alpha$, and $\sigma^2$ could be functions of redshift, and the latter two also functions of mass, we consider them to be constants for the purpose of this work. In the analytic expressions below, one may simply replace these constants with appropriate functions.  As this would introduce more degrees of freedom, and more sources of degeneracy, into the model, we defer such extensions to future work.  Our focus here is to establish a baseline model for cluster sample statistics derived with the minimum of astrophysical complications. 


\subsection{Counts, mean masses, and mass variance }\label{sec:DataVector} 

Motivated by DES-Y1 \citep{Costanzi2020DESY1} and similar analysis, we now consider two key observable quantities: i) the counts and; ii) mean masses of galaxy clusters disaggregated into bins of redshift and $\sobs$. 


Convolving the HMF, equation~\eqref{eq:hmfsimp}, with the MOR kernel, equation~\eqref{eq:MORPDF}, results in an analytic form for the space density of clusters as a function of the selection property.  While originally derived in E14, that paper did not write the form explicitly in terms of $\sobs$ but rather implicitly in terms of the mean selected mass (see equations (5), (10) and (11) of that work).  The explicit expression is 
\begin{equation}\label{eq:spaceDensity}
\begin{split}
    \ln & \left[ \frac{dn(\sobs, z)}{d\sobs} \right] = \ln A - \beta_0(z)  \\ 
& - \frac{\beta_2(z)(\sobs-\campi)^2 + 2\alpha \beta_1(z) (\sobs-\campi) -\beta^2_1(z)\sigma^2}{2(\alpha^2+\beta_2(z)\sigma^2)} ,  
\end{split}
\end{equation}
where
\begin{equation}\label{eq:A}
   A = \frac{1}{\sqrt{\alpha^2+\beta_2(z) \sigma^2}} . 
\end{equation}
The logarithmic space density is quadratic in the observable, as expected.  The last term in the second row of the expression reflects the so-called Eddington bias in mean selected mass, a topic to which we now turn.  

Following E14, Bayes' theorem implies that the mass distribution of clusters selected at fixed observable property, $\sobs$, is log-normally distributed with mean 
\begin{equation}\label{eq:mubar} 
    \langle \mu | \sobs, z \rangle =  \frac{(\sobs - \campi)/\alpha - \beta_1(z) \sigma^2/\alpha^2}{1 + \beta_2(z) \sigma^2/\alpha^2}. 
\end{equation}
The first term in the numerator is simply the inverse of the mean MOR scaling. This value is lowered by the second term, which is approximately the HMF slope, $\beta_1(z)$, times the mass variance.  
This is the same mathematics as Eddington bias, but the source of variance differs.  Eddington's scatter arose from flux measurement errors, which can be reduced by better observations.  The scatter we are dealing with here is intrinsic to the population, driven by stochastic processes within a coeval population of equal-mass halos, and so cannot be reduced by improved measurement.  We suggest \textsl{convolution bias} is a more appropriate label for this effect.

Note that the log-mean mass is actually \textit{linear} in the log-observable, $\sobs$, rather than quadratic.  This behavior arises from an exact cancellation in the quadratic terms of the Bayes' theorem derivation.  The result has an important implication ; additional information must be added to our IM analysis in order to invert the information matrix.  We take this extra constraint to be the mass variance conditioned on the observable, $s$, which is related to the MOR variance by 
\begin{equation}\label{eq:muVar} 
    \sigmamu^2(z) 
    =  \frac{\sigma^2}{\alpha^2 + \beta_2(z) \sigma^2}.
\end{equation}
Values of $\beta_2(z)$ are of order unity, so when the MOR scatter is small then the mass scatter can be approximated by the simpler expectation, $\sigmamu = \sigma/|\alpha|$.  In our IM analysis below, we impose a fractional uncertainty, $\errVarmu$, on empirical estimates of this mass variance.

Rather than log-mean mass, what is directly measured via cumulative, or \textit{stacked}, analysis of a cluster ensemble is a mean mass, with lensing mass derived from stacking weak lensing galaxy shear patterns \citep{McClintock2019DESY1lensingmass} or virial mass derived from an ensemble velocity likelihood \citep{Farahi2016EVLmass} being two viable methods.  The log of this mean mass is shifted\footnote{For a log-normally distributed $x$ having mean, $\langle x \rangle$, and standard deviation, $\sigma$, the mean of $e^x$ is $\exp[\langle x \rangle + \sigma^2/2]$.} high by $\sigmamu^2/2$
from equation~\eqref{eq:mubar}, leading to the result   
\begin{equation}\label{eq:lnMbar} 
    \ln \langle M \ | \ \sobs, z \rangle =  \frac{(\sobs - \campi)/\alpha - (\beta_1(z) -1/2) \sigma^2/\alpha^2}{1 + \beta_2(z) \sigma^2/\alpha^2}. 
\end{equation}
A systematic error floor on this quantity, $\errmubar$, is employed in the following IM analysis.


\subsection{Information Matrix Analysis}\label{sec:IMModel}

We use an information matrix approach to forecast DQ-HMF model parameter uncertainties anticipated from current and future cluster surveys.  Such analysis, while necessarily idealized, is helpful in guiding intuition and exposing parameter degeneracies.   
The observable measures we consider are traditional elements \citep{Payerne2023clusterCosmoLikelihoods} of counts and mean system masses in the observable property (\eg richness) and redshift bins as well as estimates of the mass variance at fixed property within each redshift bin.  The last two rows of Table~\ref{tab:Otherparams} list control parameters that characterize sample data quality for mean mass and mass variance measurements.

\subsubsection{Counts and mean masses}\label{IMData} 

To derive counts, $N_{s,z}$, within richness and redshift bins, we integrate the differential form, equation~\eqref{eq:spaceDensity}, in our chosen cosmology
\begin{equation}\label{eq:countsperBin} 
   N_{s,z} = \int_{z_{\rm min}}^{z_{\rm max}} dz \ \frac{dV}{dz} \ \int_{s_{\rm min}}^{s_{\rm max}} d\sobs \ \frac{dn(\sobs, z)}{d\sobs} .
\end{equation}
Here the $\{\sobs, z\}$ subscript denotes bins defined by the chosen limits of integration.  Values for these limits are given in the relevant survey application sections of \S\ref{sec:Results}.  

An exact form for the expected mean mass in each bin requires a volume-weighted integral of the exponential of equation~\eqref{eq:lnMbar}.  Motivated by the mean value theorem, we take a simpler approach by evaluating equation~\eqref{eq:lnMbar} at the median property value and midpoint redshift of each bin, 
\begin{equation}\label{eq:lnMbarperBin} 
   \ln \langle M \rangle_{s,z} = \ln \langle M \ | \ {\rm med}[\sobs], ( z_{\rm min} + z_{\rm max})/2 \rangle , 
\end{equation}
where the median value of $\sobs$ is determined by integrating the counts in each bin.


\subsubsection{Degrees of freedom}\label{IMDOF} 

We note that surveys with limited dynamic range in either redshift or selection property will be incapable of returning significant constraints on higher order terms of the DQ form.  
The cases examined in \S~\ref{sec:Results} are progressively more ambitious in terms of sample size and data quality. 

A survey limited to a single redshift shell centered on the pivot redshift, for example, returns no redshift gradient information, so the HMF parameters $\beta_{i,z}$ and $\beta_{i,z2}$ are irrelevant, leaving only the three normalizations, $\beta_{i,n}$.  These parameters, joined with the three MOR parameters, make a total of six.  

At noted above, the forms of the observable counts, equation~\eqref{eq:countsperBin}, and lensing mass measurements, equation~\eqref{eq:lnMbar}, would return only \textit{five} independent quantities: three from the quadratic counts and two from the linear log-mean mass.  A data vector consisting only of counts and mean lensing mass is thus insufficient to uniquely constrain all six model parameters.  
To produce an soluble system, we add an empirical constraint on $\sigmamu^2$, equation~\eqref{eq:muVar}.  

For optically-selected DES-Y1 clusters this quantity has been derived by \citet{Farahi2019DES} using X-ray temperatures of roughly 200 systems, finding $\sigmamu = 0.30 \pm 0.04 \ {\rm (stat)} \pm 0.09\ {\rm (sys)}$.  This estimate motivates our use of $0.3^2$ for the default MOR variance.  As improved mass estimates from lensing and dynamics become available for larger numbers of clusters, the uncertainty on this constraint is bound to improve. 


\subsubsection{Information matrix}\label{IMMatrix} 

We assume Poisson uncertainties in binned counts, a fractional error, $\errmubar$, in each mean mass measurement and a fractional error, $\errVarmu$ in the mass variance measurement.  
Using $\mathbf{p}$ to represent the set of model parameters, the information matrix for a single redshift bin takes the form 
\begin{multline}\label{eq:IMzbin}
  \mathcal{F}_{ij,z} = \sum_s \left( \frac{1}{N_{s,z}} \frac{\partial N_{s,z}}{\partial p_i} \frac{\partial N_{s,z}}{\partial p_j}  + \frac{1}{\errmubar^2} \frac{\partial \ln \langle M \rangle_{s,z}}{\partial p_i} \frac{\partial \ln \langle M \rangle _{s,z}}{\partial p_j}  \right) \\  + \frac{1}{\errVarmu^2} \frac{\partial \ln \sigma^2_{\mu}}{\partial p_i} \frac{\partial \ln \sigma^2_{\mu}}{\partial p_j}.  
\end{multline}
The first term assumes Poisson variance in the count within each observable property bin and the second assumes a constant fractional uncertainty of the mean mass measured in each bin.  The final term accounts for uncertainty in the mass variance, which is taken to be property-independent but depends on redshift through the $\beta_2(z)$ term in equation~\eqref{eq:muVar}.  We evaluate this term at the midpoint of the redshift bin under consideration.  

For the survey-specific expectations, the full information matrix is determined by a sum over all redshift bins 
\begin{equation}\label{eq:IMfull}
    \mathcal{F}_{ij} \ = \ \sum_z \  \mathcal{F}_{ij,z}. 
\end{equation}
For the DES-Y1 case, the three redshift bins used in the \citet{Costanzi2020DESY1} analysis,  $z \in [0.2, 0.35)$, $[0.35, 0.5)$ and $[0.5, 0.65)$ are employed.  For LSST, we use seven equally spaced redshift bins covering the interval $0.1 < z < 1.5$.  

Appendix~\ref{sec:ToyModel} provides explicit analysis of a reduced toy model based on a single redshift and property bin and having only three free parameters.  This example helps illustrate the coupling of HMF and MOR parameters but the simplified scenario (two of the three MOR parameters are known) limits generalization of the results to the more complex survay applications. 

\section{HMF Parameter Forecasts}\label{sec:Results} 

We now explore potential parameter constraints from current and future optical cluster surveys under two conditions for the quality of derived mean mass per bin and mass variance.  We refer to these conditions as Weak and Strong, with the latter having a factor two or better levels of uncertainty compared to the former.  

The choices of fractional errors in mean mass per bin, $\errmubar$, and mass variance, $\errVarmu$, are summarized in Table~\ref{tab:QualityCases}.  The Weak choices for DES-Y1 of $(0.1, 0.6)$ are based on current systematic uncertainties estimates  \citep{McClintock2019DESY1lensingmass, Farahi2019DES}, while the Strong assumption improves each by a factor of two.  The LSST Weak quality are slightly improved over DES-Y1 Strong, and the LSST Strong values represent a further improvement of a factor four, to $\errmubar = 0.01$ and $\errVarmu = 0.05$.  

The latter constraints are certainly aspirational and will require substantial effort to achieve.  For example, the fractional error in mass scatter, $\Delta \sigmamu / \sigmamu = \errVarmu /2$, is only 2.5\% in the LSST Strong case, an uncertainty of only $0.0075$ on a central value of $0.300$.  For the LSST Weak case, the uncertainty in mass scatter would be a less stringent value of $0.015$.   


In common with much IM analysis, the spirit of this work is to reveal best-case DQ-HMF+MOR parameter constraints from analysis of counts, mean mass and mass variance assuming no prior knowledge.  Our example applications are tuned to optical cluster surveys with galaxy richness, $\lambda$, as the observable selection property but the model can be generalized to searches at other wavelengths.  For example, on the mass scales investigated here, hydrodynamic simulations suggest that hot gas mass has a smaller intrinsic variance, $\sim \! 0.1^2$, at the pivot mass scale for redshifts, $z<1$ \citep{Farahi2018BMscaling}. Benefits of a sharper proxy are discussed in \S~\ref{sec:discSelectionML}.

\begin{table}
    \centering
\begin{tabularx}{0.7\columnwidth}{l | c c | c c }
     &  \multicolumn{2}{c|}{DES-Y1} & \multicolumn{2}{c}{LSST}  \\
    Level      & $\errmubar$  & $\errVarmu$  & $\errmubar$  & $\errVarmu$ \\
        & & & & \\      [-8 truept]    
     \hline \\[-6 truept]
    Weak & 0.10 & 0.60  & 0.04  & 0.20 \\
    Strong & 0.05 & 0.30  & 0.01  & 0.05 
\end{tabularx}
    \caption{Assumed fractional errors mean mass and mass variance for the IM analysis, equation~\ref{eq:IMzbin}.  For each sample, two levels of quality, Weak and Strong, are used, with the latter improving over the former by a factor of two (DES-Y1) or four (LSST).}
    \label{tab:QualityCases}
\end{table}

\subsection{Current survey application: DES-Y1}\label{sec:DESY1}

The DES-Y1 survey identified 6500 optical clusters with $\lambda \ge 20$ lying at redshifts $0.2 < z < 0.65$ within roughly 5000 square degrees of the southern sky using the redMaPPer algorithm \citep{Rykoff2016DES-SVredmapper}.  For each of the three redshift bins, counts and mean weak lensing masses within four richness bins of 
$\lambda \in [20,30)$, $[30,45)$, $[45,60)$ and $\ge 60$ 
were determined by \citet{Costanzi2020DESY1} 
 and \citet{McClintock2019DESY1lensingmass}, respectively.   
The DES-Y1 cluster counts above a richness of 20 in the $[0.2,0.35)$, $[0.35,0.5)$ and $[0.5,0.65)$ redshift ranges were 1352, 2556 and 2596, respectively.  
Our reference model, which uses a different cosmology and MOR, yields a similar total count but with slightly different counts per redshift bin (1498, 2286, and 2752), 
a level of agreement acceptable for the purpose of illustration. 
We employ the same four richness bins as DES-Y1 in the IM analysis.  

Due to the limited redshift range in the DES-Y1 sample, we ignore the highest-order terms from the $\beta_i$ redshift expansions, equations~\eqref{eq:betaz} and \eqref{eq:betatwoz}.  
The model thus has eight degrees of freedom consisting of five HMF and three MOR parameters.

\begin{figure}
    \centering
    \includegraphics[width = 1.0\columnwidth]{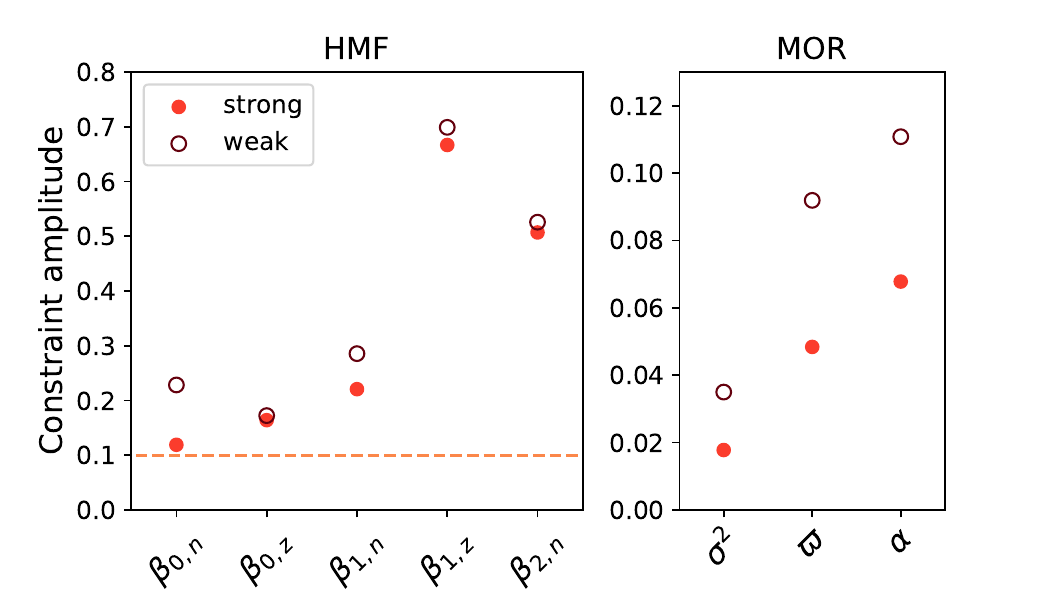}
    \caption{IM-forecasted parameter uncertainties for a DES-Y1-like cluster sample under the data quality cases listed in Table~\ref{tab:QualityCases}.  A reduced DQ-HMF model based on the five lowest-order terms (left panel) is employed.  The orange line indicates an 0.1 reference value, reproduced in Figure~\ref{fig:LSSTdiagonal}. The right panel shows MOR parameter forecasts.  Informative priors on these could potentially reduce HMF parameter uncertainties.  Parameter correlations are shown in Figure~\ref{fig:DEScorner}.  
    }
    \label{fig:DESdiagonal}
\end{figure}

\begin{table}
	\centering
 \begin{tabularx}{0.5\columnwidth}{c | c  |  c }
   &  \multicolumn{2}{c}{Quality}  \\   
   Parameter & Weak & Strong \\ 
     & &  \\      [-8 truept]    
 \hline\\[-6 truept]
 $\beta_{0,n}$  & 0.23 & 0.12 \\ 
 $\beta_{0,z}$ & 0.17 & 0.16 \\ [2 truept]    
 $\beta_{1, n}$ & 0.29 & 0.22 \\
 $\beta_{1, z}$ & 0.70 & 0.67 \\ [2 truept]    
 $\beta_{2, n}$ & 0.53 & 0.51 \\ [2 truept] 
 $\sigma^2$ & 0.035 & 0.018 \\ 
 $\campi$ & 0.092 & 0.048 \\
 $\alpha$ & 0.11 & 0.068 \\ 
\end{tabularx}
 \caption{DQ-HMF and MOR parameter constraints anticipated from a sample patterned after $\lambda > 20$  DES-Y1 clusters, shown in Figure~\ref{fig:DESdiagonal}.}
 \label{tab:DESparams} 
\end{table}

Applying the IM analysis using the counts and mean masses in these twelve bins, along with the uncertainty in mass variance within each redshift bin, yields the parameter constraints listed in Table~\ref{tab:DESparams} for the Weak or Strong quality assumptions.  Figure~\ref{fig:DESdiagonal} plots these parameter uncertainties, with the orange line offering a reference value of 0.1 for future reference to LSST sample expectations. 

Under the Weak quality case, the normalization at the pivot mass and redshift, $\beta_{0,n}$, is forecast to have an uncertainty of $0.23$, implying a fractional uncertainty of $26\%$ in the number density.  The redshift gradient of the normalization, $\beta_{0,z}$, is slightly better constrained, at 0.17, which represents a $7\%$ fractional uncertainty on a central value of $2.38$ in \lcdm.  This result is helped by the fact that the MOR is independent of redshift; the change in counts across redshift in each richness bin feeds information primarily to $\beta_{0,z}$. 

The mass slope of the HMF at the pivot redshift, $\beta_{1,n}$, is forecast to have an uncertainty of $0.29$, which is $14\%$ of its \lcdm\ central value of $2.38$.  The highest order terms, those describing the redshift gradient of the slope with mass, $\beta_{1,z}$, and the mass curvature at $z_p$, $\beta_{2,n}$, are forecast to be weakly constrained with errors $>0.5$ on central values of $1.33$ and $0.75$, respectively.  

Under the Weak quality case, forecast errors on MOR normalization and slope are $\sim 0.1$.  The intrinsic variance of the observable conditioned on mass, $\sigma^2$ is anticipated to be returned within an error of $0.035$ on a central value of $0.09$.  Note that these constraints come entirely from the sample itself.  In practice, one might imagine external priors on these parameters being imposed in informative ways.  

\begin{figure*}
    \centering
    \includegraphics[width = 1.6\columnwidth]{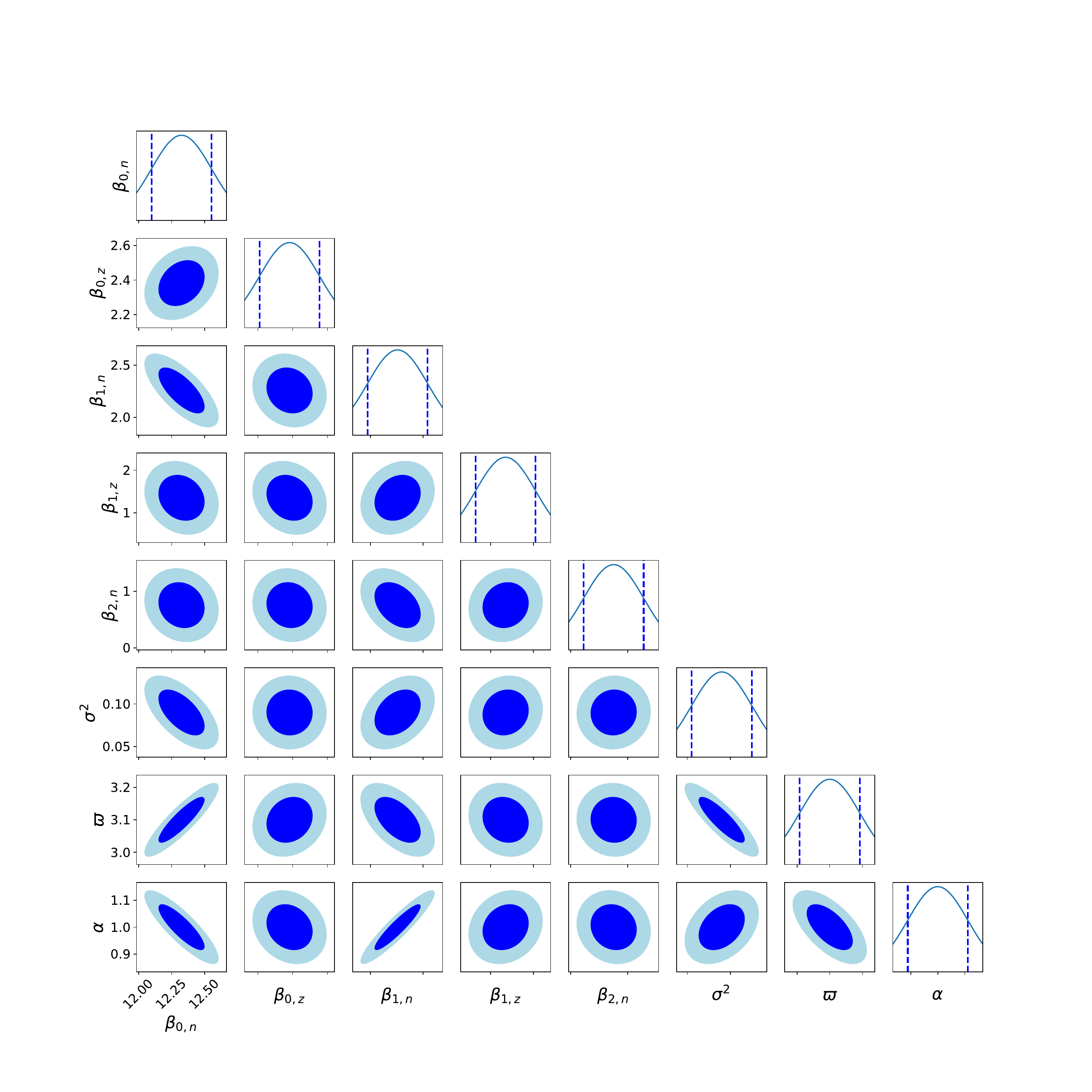}
    \vspace{-24truept} 
    \caption{Parameter covariance forecast for a survey patterned after DES-Y1 under Weak quality assumptions (see Table~\ref{tab:QualityCases}).
    }
    \label{fig:DEScorner}
\end{figure*}

The couplings between parameters for the Weak quality case are displayed in Figure~\ref{fig:DEScorner}.  From the analytic expressions for the space density, equation~\eqref{eq:spaceDensity}, and log-mean mass, equation~\eqref{eq:mubar}, we can anticipate significant degeneracies among MOR and HMF parameters.  Unsurprisingly, the two normalization parameters, $\beta_{0,n}$ and $\campi$, are strongly coupled, as are the two slope measures, $\beta_{1,n}$ and $\alpha$.  The MOR intrinsic variance, $\sigma^2$, couples strongly to all of these parameters.  

The fact that the DES-Y1 richness threshold of 20 lies close to the MOR normalization, $e^\campi = 22$, means that the limiting mass scale lies close to the pivot mass $M_p$.  Because counts and mean masses in higher richness bins provide leverage to only one side of $M_p$, there is non-zero covariance between the HMF normalization, slope and curvature at the pivot redshift.  These correlations are somewhat weaker than those associated with the MOR.  Because the MOR is assumed to be a pure power-law, with zero curvature, there is very weak coupling between MOR parameters and the HMF curvature, $\beta_{2,n}$.  

Forecasts for the Strong quality case
are shown as filled circles in Figure~\ref{fig:DESdiagonal}.  The MOR sector receives the primary benefit of improved quality in mean mass and mass variance, with improvements close to a factor of two.  In the HMF sector, the pivot normalization and mass slope, $\beta_{0,n}$ and $\beta_{1,n}$, see significant improvement while the remaining terms improve only modestly.  
Because the counts in each bin are the same for the two cases, the redshift gradient parameters,  $\beta_{0,z}$ and $\beta_{1,z}$, are little improved. 

A picture that emerges is that improved measurements of mean mass and mass variance tighten the MOR sector, and these improvements filter primarily into the HMF pivot normalization and slope and secondarily to higher-order HMF parameters.  This behavior is repeated in the LSST analysis below. 

Note that the forecast uncertainty in HMF pivot normalization does not include any contribution from errors in distance measurements.  At the forecast level of 0.12 for the Strong quality case, current volume uncertainties are sub-dominant, though still contribute at the level of $\sim \! 0.08$ \citep{SDSSIII2017BAOdistance, DES2022BAOdistance}.  


\subsection{Future survey application: LSST}\label{sec:LSST}

The Rubin Observatory Legacy Survey of Space and Time (LSST) will be both wider and deeper than DES \citep{2019LSSTreferenceDesign, Chisari2019ApJS_LSSTcode}.  The increase in depth will yield improved measurements of galaxy shapes and colors, and this improvement should translate to more precise estimates of weak lensing mass.  For the quality of mean mass estimates, we employ systematic error levels of 0.04 (Weak) and 0.01 (Strong).  For the quality level of mass variance we assume values of 0.20 and 0.05, respectively.  The LSST Weak values represent modest improvements over the DES Strong case. 

To push this idealized case further, we anticipate that improvements in optical cluster finding will allow for a factor of two reduction in the sample richness limit, to a value of 10.  The overall number of clusters expected using this observable threshold is 380,000, and the redshift distribution of the counts is similar to that shown for the mass-limited case of Figure~\ref{fig:zbinCounts}. Note, however, that in the IM analysis we use seven redshift bins, each of width 0.2, between $0.1<z\le 1.5$, as well as five richness bins comprised of the four DES-Y1 bins joined with $\lambda \in [10,20)$.  The total number of terms in the IM is 77, consisting of 35 counts, 35 mean masses, and seven mass variance measures.  


\begin{figure}
    \centering
    \includegraphics[width = 1.0\columnwidth]{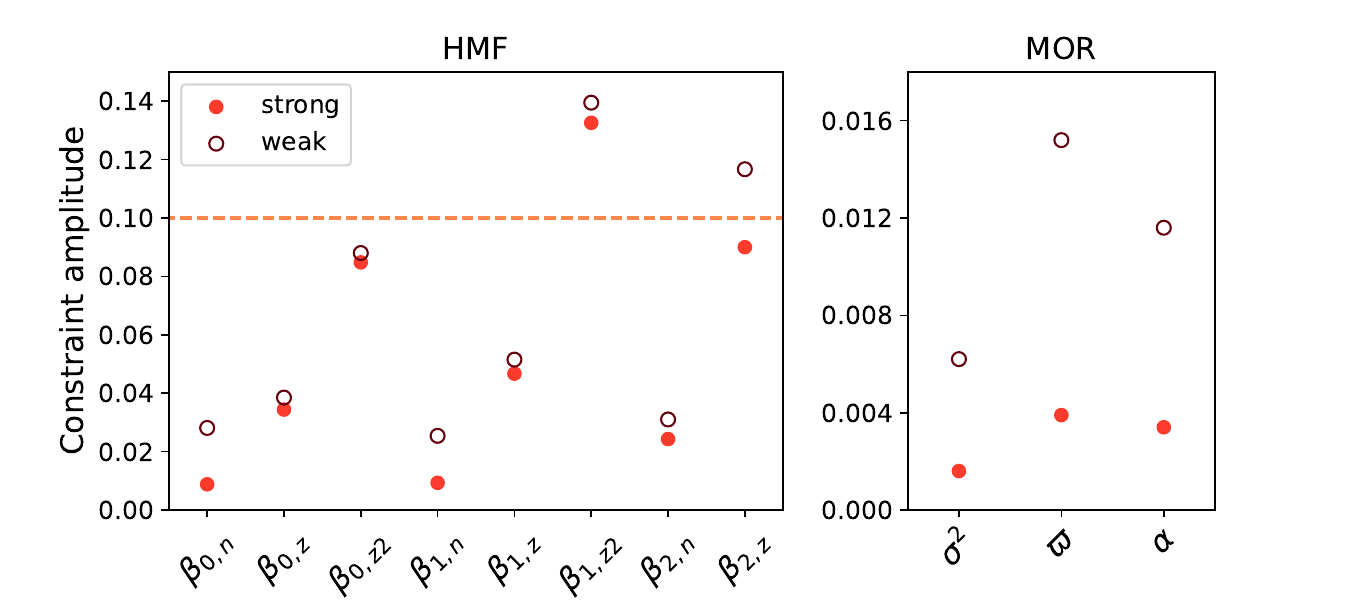}
    \caption{IM-forecasted uncertainties on DQ-HMF (left) and MOR (right) parameters from an LSST-like optical survey of  $\lambda > 10$ clusters under the quality assumptions listed in Table~\ref{tab:QualityCases}.  Five richness bins and seven redshift bins across $0.1<z<1.5$ are employed. The reference value of 0.1 is repeated from Figure~\ref{fig:DESdiagonal}.  Parameter correlations are displayed in Figure~\ref{fig:LSSTcorner}. }
    \label{fig:LSSTdiagonal}
\end{figure}

\begin{table}
	\centering
 \begin{tabularx}{0.5\columnwidth}{c | c  |  c }
   &  \multicolumn{2}{c}{Quality}  \\   
   Parameter & Weak & Strong \\ 
     & &  \\      [-8 truept]    
 \hline\\[-6 truept]
 $\beta_{0,n}$  & 0.028 & 0.0088 \\ 
 $\beta_{0,z}$ & 0.039 & 0.034 \\ 
 $\beta_{0,z2}$ & 0.088 & 0.085 \\  [2 truept]    
 $\beta_{1, n}$ & 0.025 & 0.0093 \\
 $\beta_{1, z}$ & 0.052 & 0.047 \\
 $\beta_{1, z2}$ & 0.14 & 0.13 \\  [2 truept]    
 $\beta_{2, n}$ & 0.031 & 0.024 \\
 $\beta_{2, z}$ & 0.12 & 0.090 \\ [2 truept]
 $\sigma^2$ & 0.0062 & 0.0016 \\ 
 $\campi$ & 0.015 & 0.0039 \\
 $\alpha$ & 0.012 & 0.0034 \\ 
\end{tabularx}
 \caption{DQ-HMF  and MOR parameter constraints anticipated from a sample patterned after $\lambda > 10$ LSST clusters, shown in Figure~\ref{fig:LSSTdiagonal}. }
 \label{tab:LSSTparams} 
\end{table}
 
The larger counts and better quality assumptions increase the volume of the IM matrix determinant relative to the DES-Y1 case (see the toy model in Appendix \ref{sec:ToyModel}).  We thus employ the full set of eight DQ-HMF parameters.  

Figure~\ref{fig:LSSTdiagonal} and listed in Table~\ref{tab:LSSTparams} show that, despite the increase in model dimension, the larger information content improves the constraints on all parameters relative to DES-Y1. In the left panel of Figure~\ref{fig:LSSTdiagonal}, the horizontal dashed line reproduces the 0.1 amplitude in Figure~\ref{fig:DESdiagonal}.  While none of the forecast uncertainties fall below this value for DES-Y1, in the LSST-Weak case all but two high-order parameters, $\beta_{1,z2}$ and $\beta_{2,z}$, lie below it.  

In the LSST-Weak case, all three HMF shape parameters at the pivot redshift, $\beta_{0,n}, \, \beta_{1,n}$ and $\beta_{2,n}$ are forecast to have uncertainties of 4 percent or better.  For the Strong quality case, the pivot normalization and mass slope have forecast errors of one percent.  

Tight constraints on the three MOR parameters emergy, shown in the right panel of Figure~\ref{fig:LSSTdiagonal}.  For the Weak quality case, the MOR normalization and slope are forecast to have uncertainties of 0.015 and 0.012, respectively.  In the Strong case, the forecast errors below 0.004 for both.  Such sub-percent errors reflect the powerful potential of LSST-era samples, but achieving such tight constraints will be difficult in practice, as discussed in \S\ref{sec:Discussion} below. 

The property variance uncertainties translate to errors on the richness scatter (square root of variance), of $\pm 0.01$ and  $\pm 0.003$, in the Weak and Strong cases, respectively.  These values, roughly three and one percent fractional errors on the central value $\sigma = 0.3$, will again be quite challenging to achieve in practice.  


\begin{figure*}
    \centering
    \includegraphics[width = 2.2\columnwidth]{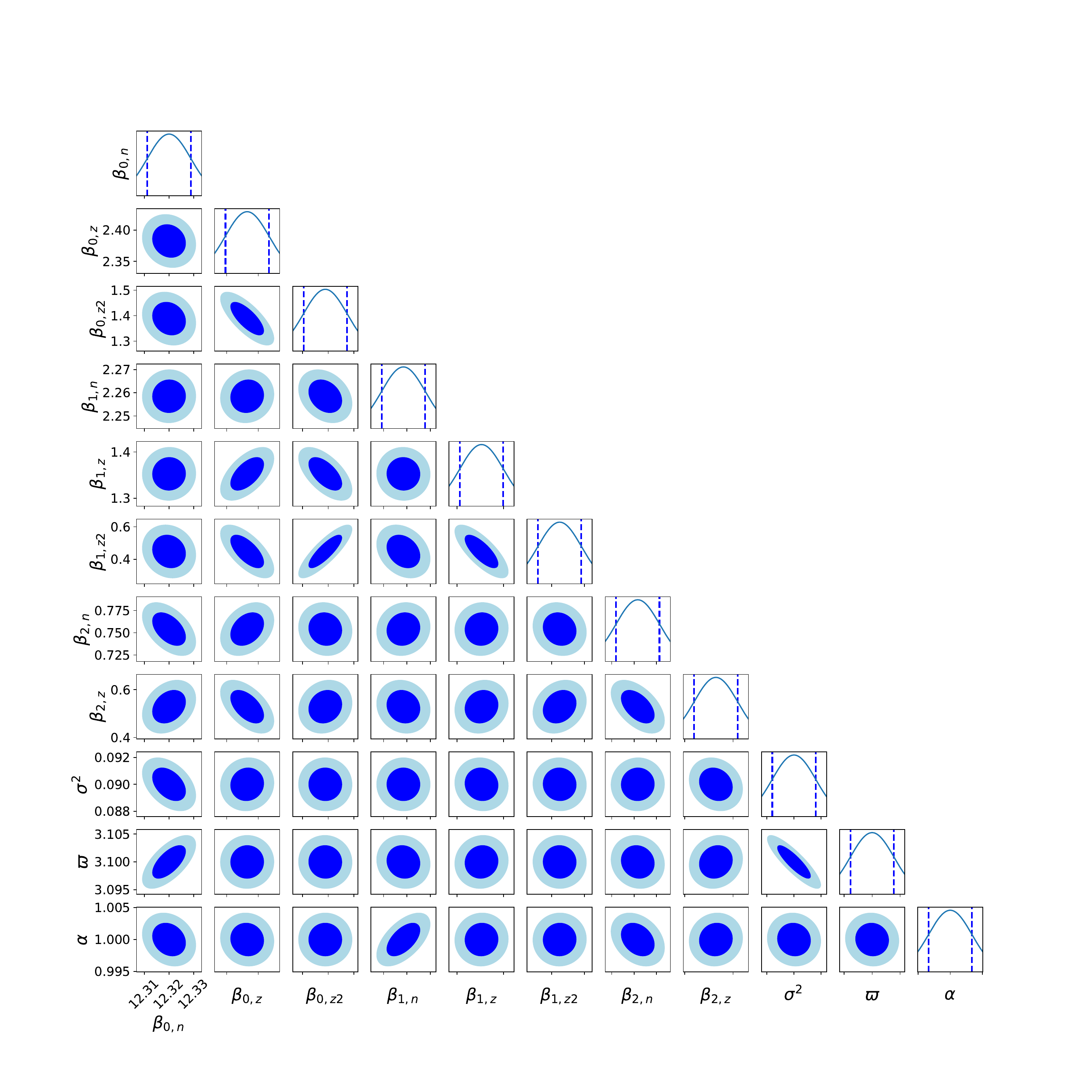}
       \vspace{-40truept} 
 \caption{Parameter covariance for an LSST-like survey under Strong quality constraints for weak lensing mass and mass variance measurements.
    }
    \label{fig:LSSTcorner}
\end{figure*}

The full IM covariance structure under Strong data quality assumptions for LSST is shown in Figure~\ref{fig:LSSTcorner}.  As in the DES-Y1 example, the MOR variance and normalization parameters ($\sigma^2$ and $\campi$) remain strongly coupled, as are the MOR and HMF normalizations ($\campi$ and $\beta_{0,n}$) and slopes ($\alpha$ and $\beta_{1,n}$).  

For the LSST case, our choice of pivot redshift, $z_p = 0.5$, lies below both the median sample redshift and the redshift midpoint of 0.8.  As a result, redshift evolution parameters of the DQ-HMF are coupled.  For example, the redshift gradient of the HMF normalization, $\beta_{0,z}$, is mildly correlated with its redshift curvature, $\beta_{0,z2}$, and the redshift curvature of the local slope, $\beta_{1,z2}$.  While a more optimal choice of pivot redshift would reduce these correlations, we maintain a common pivot redshift for both samples analyzed here in order to provide a fair comparison of potential gains.  

Relative to the DES-Y1 analysis, the lower richness threshold assumed for the case of LSST  offers leverage below the pivot mass scale of $10^{14.3} \hinv \msol$.  Unlike the DES-Y1 case, the HMF shape parameters at the pivot redshift ($\beta_{0,n}, \, \beta_{1,n}, \, \beta_{2,n}$) are largely uncorrelated for the case of LSST.  

Increasing the richness limit to 20, the overall sample size drops to 60,000, a factor of roughly $2^{2.5}$ lower than the richness 10 counts.  Parameter constraints are degraded accordingly, with forecast errors in the Strong case of $0.04$ in $\beta_{0,n}$ and $0.08$ in $\beta_{1,n}$ and $\beta_{1,n}$. 


\section{Discussion}\label{sec:Discussion}


In \S~\ref{sec:MiraTitanFits} we showed that a compact form sufficiently captures the near-field, space-time density of high mass halos derived from an N-body emulator, where sufficiency here is in relation to current systematic uncertainties associated with the effects of galaxy formation feedback. The eight DQ-HMF parameters have straightforward interpretations as polynomial  coefficients in log-mass and redshift.  

A key benefit of the model is that convolution with a log-normal MOR produces closed-form expressions for observable features of group and cluster samples: counts, mean mass, and mass variance as a function of an observable property and redshift.  Information matrix analysis designed around existing and planned optical cluster surveys indicate that potential constraints from an LSST-scale survey could be percent-level on many DQ-HMF and MOR parameters.  

Achieving such precise constraints will be challenging.  We begin by discussing the role of volume uncertainties, projection effects and more complex MOR forms. We then briefly touch on sample selection, focusing on the potential benefits of joining multi-wavelength samples.  In the case of LSST, we show that constraints on all model parameters could be improved using a one-tenth subsample of clusters having a tighter mass proxy, with intrinsic property variance of $0.1^2$ rather than $0.3^2$. Machine learning techniques that employ all available measurements could provide a pathway to classifying such a sample, particularly if tuned accurately by synthetic data from cosmological hydrodynamics simulations.

\subsection{Comoving volume uncertainties } \label{sec:discVolume}

In the IM forecasts above, we have ignored uncertainties in the comoving 
comoving cosmic volume.  
Uncertainties in cosmic volume will introduce additional error to the normalization terms,  
$\beta_{0,x} \in \{ \beta_{0,n},~ \beta_{0,z},~ \beta_{0,z2} \}$.  
At the chosen pivot redshift of 0.5, SDSS-III measurements of baryon acoustic oscillation (BAO) and galaxy clustering \citep{SDSSIII2017BAOdistance} constrain the local cosmic volume to within $4\%$, and the $2.7\%$ distance measurement to $z=0.835$ from DES BAO analysis \citep{DES2022BAOdistance} implies a roughly 8\% error in volume.  These uncertainties are subdominant to the DES-Y1 IM errors in $\beta_{0,x}$ (Table~\ref{tab:DESparams}) but achieving future LSST constraints (Table~\ref{tab:LSSTparams}) will require more precise distance measurements.  

Increased precision will almost certainly come.  For example, current forecasts for distances derived from Type Ia SN in LSST suggest comoving distance errors of roughly $0.25\%$ in the redshift range $0.5 < z < 1.2$ \citep{Mitra2023LSST_SNdistance}, meaning volume errors below one percent. This is similar to the $1\%$ LSST-Strong constraint on $\beta_{0,n}$ from the IM analysis.  
Alternatively, the HMF could simply be redefined in terms of the directly observable volume element, in units of number per square degree per unit redshift rather than cubic megaparsecs.  The above IM forecasts apply directly to this alternative HMF framing. 

\subsection{Projection and more complex MOR forms } \label{sec:discMOR}

A galaxy cluster sample defines a discrete population in which each member is minimally defined by a location on the sky and (ideally) redshift, an angular size, and one or more aggregate observable properties, preferably measured within that aperture. The inverse mapping of a given cluster sample onto the underlying space-time population of massive halos is complicated by several effects arising from projection and other factors such as halo orientation.  In addition, the minimal MOR used above may require extensions for practical application to specific cluster surveys. 

These non-trivial issues pose a challenge to precise modeling of the cluster--halo connection, especially for optically-selected samples \citep[\eg][]{Wetzell2022DESY3veldisp, Zhou2023intrinsicAlignDESY1, Varga2022syntheticClustersDESY3, Giles2022XCS+SDSSRM, Upsdell2023XCS+DES, Zhang2023clusterTriaxialityDES}.  We sketch here some ideas for how to incorporate them into DQ-HMF-focused analysis. 

\subsubsection{Projection}\label{sec:projection} 
Typically, a single massive halo subtends a few arcminutes of sky and, due to its origin as a peak in an initially Gaussian noise field \citep{Kaiser1984}, tend to be more strongly clustered than the general dark matter distribution.  
The intrinsic properties of a given halo are thus superposed with projected contributions from other halos along the same line-of-sight.   A general way to accommodate this effect on measured properties is by adding another statistical factor,  $p(\Smeas | S, M, z)$, that accounts for projection-induced distortions \citep[\eg][]{Mulroy2019LoCuSS}.  This function will introduce additional parameters, prior values of which can be estimated by survey-specific simulations \citep{Costanzi2019opticalProjection, Chiu2020CAMIRA_MOR, LSSTDESC2021DC2}.  

Projection will generally boost aperture-based signals \citep{White2002SZsims, Cohn2007MillenRSclusters, Costanzi2019opticalProjection}, driving positive skewness into the observed property kernel, $P(s_{\rm obs} | \mu, z)$.  The inverse kernel, $P( \mu |s_{\rm obs}, z)$, will lean toward lower halo masses, and this implies a similar lean in potential well depth measures such as X-ray temperature \citep{Ge2019maxBCGextreme}.   

In terms of the model, kernel skew can be accommodated multiple ways, including by a Gaussian mixture
\begin{equation}\label{eq:mixMod} 
    P(\sobs | \mu) = f \mathcal{N}(\overline{s}_{\rm obs}(\mu), \sigma^2) + (1-f) \mathcal{N}((\overline{s}_{\rm obs}(\mu) +\Delta_p), \sigma_p^2) , 
\end{equation}
where the first term represents a majority fraction, $f$, of clear sightlines with mean $\overline{s}_{\rm obs}(\mu)$ and variance $\sigma^2$, and the second term represents a highly-projected subset boosted in the mean by $\Delta_p$ with variance $\sigma_p^2$.  This form, which is supported by red sequence cluster finding using Millennium Simulation galaxies \citep{Cohn2007MillenRSclusters}, 
brings the benefit of retaining the analytical forms in \S\ref{sec:DataVector} which would be fast to compute in survey 
analysis.  A downside is the introduction of three additional parameters, but these dimensions could be coupled and reduced to a simple skewness measure implemented by Markov Chain Monte Carlo chains. 



\subsubsection{Intrinsic MOR Complexity}\label{sec:MOR} 

Intrinsic property statistics are sensitive to both cosmology (through environmentally-sensitive formation histories) and astrophysics related to galaxy formation and plasma evolution.  The minimal MOR form used above, with three parameters, is likely to require some extensions for precise survey likelihood application.  Based largely on the behavior of halos in cosmological hydrodynamics simulations, we briefly outline modifications that may apply to different observable properties.  

\emph{MOR shapes from cosmological hydrodynamics simulations.}  Large samples of high-mass halos from cosmological hydrodynamics simulations provide the means to test the MOR kernel for multiple observable properties.  In BAHAMAS+MACSIS simulations, the hot gas mass and the total stellar mass within $\rtwoh$ follow log-normal kernel shapes \citep[][hereafter F18]{Farahi2018BMscaling}.  

The existence of a log-normal PDF for the total stellar mass of halos was confirmed using three independent cosmological hydrodynamics simulations by \citet{Anbajagane2020stellarProps}.  
That work also finds slight skewness in halo mass-conditioned statistics for the total number of satellite galaxies, $\Nsat$, and the BCG stellar mass, $\mstarBCG$.  A common Gaussian mixture fit  is derived for the normalized $\Nsat$ kernel, with $79 \pm 1$ percent of halos in a dominant component with mean, $0.28 \pm 0.01$, and scatter, $0.68 \pm 0.01$, and the remaining 21\% component having mean $-1.04 \pm 0.05$ and scatter $1.13 \pm 0.02$.  
More work is needed to understand intrinsic MOR shapes for other observable properties, such as X-ray luminosity and temperature or thermal SZ decrement amplitude, and efforts to verify statistic forms from different cosmological hydrodynamics methods are also warranted.

\emph{Running of MOR parameters with redshift and/or halo mass.}  The property normalization, $\campi$, is likely to evolve with redshift.  A self-similarity assumption \citep{Kaiser1986SS} that ties physical properties to the evolving critical density is often used to express, $\campi(z)$, in terms of powers of $E(z) \equiv H(z)/H_0$.\footnote{This form is appropriate for the critically-thresholded $\mtwoh$ halo mass convention employed here; using mean mass rather than critical density in the spherical overdensity condition leads to powers of $1+z$ instead of $E(z)$.} Under strict self-similarity, the total stellar or gas mass fractions are independent of redshift.  In the BAHAMAS+MACSIS simulations, F18 find modest (several percent) redshift dependence in both measures, with the gas mass fraction declining, and stellar mass fraction increasing, slightly from $z=1$ to $z=0$.  These shifts are mildly mass-dependent, being larger at lower halo masses that are more strongly influenced by galaxy evolution. 
Free parameters introduced to capture deviations in normalization from self-similarity would couple most strongly to the DQ-HMF normalization parameters, $\beta_{0,x}$.  The intrinsic property variance, $\sigma^2$, of hot gas and stellar mass was also found to run weakly with mass and redshift by F18.  

The constancy of the MOR slope, $\alpha$, is also a simplification that may require modification for some properties.  For example, F18 find that the slopes of hot gas mass and stellar mass vary modestly with both halo mass scale and, for the former, redshift.  At lower halo masses, the hot gas mass slope steepens to values above unity, and the stellar mass scaling becomes shallower than unity.  A parameter introduced to describe an MOR slope gradient, $d \alpha / d \mu$, would couple most strongly to the MOR curvature, $\beta_{2,n}$.   Extending further to allow for this parameter to run linearly with $(1+z)$ would then couple to $\beta_{2,z}$.

\subsection{``Gold Sample'' Selection with Machine Learning using Multiple Properties}\label{sec:discSelectionML}

Cluster samples are generally defined by a threshold in a single observed selection property. The DES-Y1 sample, for example, is limited by red galaxy richness, $\lambda \ge 20$.  
The mapping between a set of observed clusters and their underlying host halos is assumed to be bijective; a chosen halo maps uniquely to a single cluster, and vice-versa.  This is not always the case\footnote{See the spectacular case of Planck Sunyaev-Zeldovich source PSZ1 510, which represents a near perfect alignment on the sky of two rich ($\lambda \sim 80$) clusters offset by 0.1 in redshift \citep{Rozo2015redMaPPerIII}.}, 
and multi-wavelength studies are critical to understanding how frequently this assumption is violated.  In a recent joint study of cluster samples identified independently by X-ray and optical observations in roughly 60 deg$^2$ of sky, \citet{Upsdell2023XCS+DES} find that only one of 178 X-ray sources has two optical clusters identified along the same line of sight.  Such effects, as well as more prosaic issues such as survey masking \citep[\eg][]{Rykoff2016DES-SVredmapper}, will affect cluster selection and require calibration by multi-wavelength observations and simulations. 

Joint property analysis of large cluster samples can improve cosmological parameter constraints \citep{Cunha2009clusterCrossCorrelation} because combining multiple observable properties can substantially reduce mass variance relative to single-property characterization \citet{Ho2023MLmasses}. The anti-correlation of hot gas and stellar mass contents observed in the LoCuSS sample \citep{Farahi2019LoCuSS} is an important feature; selecting on just these two intrinsic properties in the Magneticum simulation yields a variance in halo mass of $0.05^2$ \citep{Ho2023MLmasses}.

\subsubsection{Potential Gains of a Gold Sample} \label{sec:goldSample}

There is potential to improve DQ-HMF parameter constraints using a selection approach that identifies a Gold Sample of clusters with reduced intrinsic MOR variance.  For this example, we imagine a classifier returning 10\% of the overall population with intrinsic MOR variance, $0.1^2$.  While a significant improvement over the $0.3^2$ value used in our default analysis, we note that, for high halos masses, the hot gas mass is seen to have such a small variance \citep{Truong2018XrayScaling, Farahi2018BMscaling, Pop2022TNGscaling, Farahi2022,  Pellissier2023RhapsodyC}.

\begin{figure}
    \centering
    \includegraphics[width = 0.95\columnwidth]{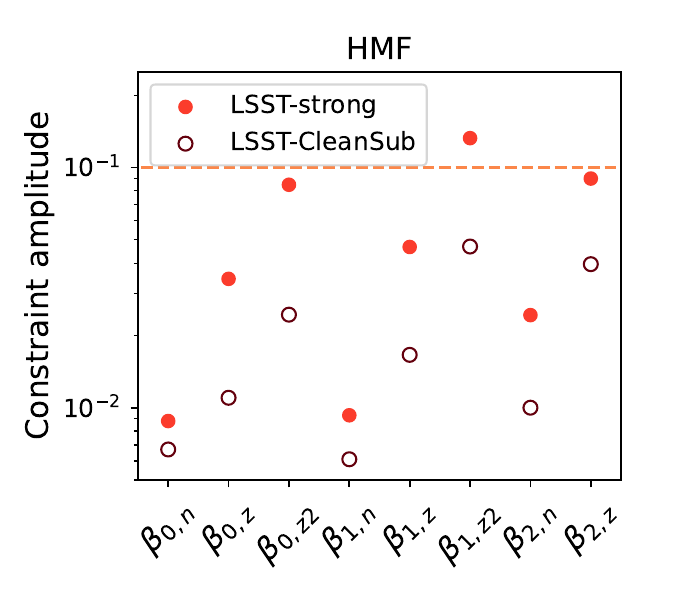}
       \vspace{-8truept} 
 \caption{HMF parameter constraints for the LSST-Strong case with MOR variance, $\sigma^2 = 0.3^2$ (filled circles, same as Figure~\ref{fig:LSSTdiagonal}), are compared to those from a cleaner subset (``Gold Sample'') consisting of 10\% of the former sample with a reduced MOR variance of $0.1^2$ (open circles).  The clean subset yields improvements, particularly in the higher-order quantities such as $\beta_{0,z2}$, $\beta_{1,z2}$, and $\beta_{2,z}$.  Note the logarithmic scale on the constraint amplitude.  
    }
    \label{fig:CleanSub}
\end{figure}

Using the reduced, three-parameter model of Appendix~\ref{sec:ToyModel} as a guide, the information volume scaling of $N \sigma^{-4}$ for low-scatter proxies (other parameters held fixed), equation~\eqref{eq:FsimpleDet}, would imply that the improvement in MOR variance wins over the decrease in sample size. 
Figure~\ref{fig:CleanSub} confirms this to be the case. The filtered cluster subsample with 10 percent of the counts but $0.1^2$ variance yields improvements in all HMF parameters, with the biggest gains occurring for the highest order quantities, $\beta_{0,z2}$, $\beta_{1,z2}$ and $\beta_{2,z}$.  As discussed in \S\ref{sec:neutrinos}, the shifts in such higher-order terms caused by massive neutrinos are of the order 0.01, potentially within reach of Gold Sample analysis.  



Machine learning (ML) techniques have been demonstrated to yield improved estimates of galaxy cluster masses from noisy observations derived from simulations of massive halos \citep{Ntampaka2016MLdynamicalMass, Ntampaka2019MLmasses, CohnBattaglia2020MLmasses, Krippendorf2023ERositaMLmasses, Ho2023MLmasses}, and sample selection in the low signal-to-noise regime has been explored by \citet{Kosiba2020}. Symbolic regression has been used to identify property combinations that minimize mass variance \citep{Wadekar2023} and random forest techniques have been used to classify galaxies into orbit classes using projected phase space information \citep{Aung2023MLgalaxyOrbits, Farid2022MLgalaxyOrbits}.  


We encourage other researchers to explore whether ML methods can be trained to identify a Gold Sample with characteristics similar to that assumed above.  Synthetic sky maps and catalogs are essential elements for such studies, and more effort is needed to move beyond single wavelength products \citep{DeRose2019Buzzard, LSSTDESC2021DC2, Wechsler2022ADDGALS, Kovacs2022validation, Frontiere2022Farpoint, Troxel2023} toward synthetic lightcone products with joint stellar, gas, and dark matter properties \citep{Omori2022AgoraSim, Osato2023BaryonPasting, Schaye2023Flamingo}.  Deep learning methods could facilitate production of such maps \citep{Han2021}.
As multiple synthetic skies that jointly meet the requirements of surveys in optical/IR, sub-millimeter and X-ray become available, methods for sample selection can be cross-verified,  
trained on one simulation methodology and tested on another.

\subsection{Lensing and Correlated LSS Measures} \label{sec:discLensing}

Massive halos impose peaks in weak lensing maps on arcminute scales, and tangential shear analysis has long been a staple method of estimating the underlying true halo masses of galaxy clusters \citep{TysonValdesWenk1990lensing, MiraldaEscude1991lensing,  KaiserSquires1993lensing, LuppinoKaiser1997MS1054lensing}, see the review of \citet{Hoekstra2013lensingReview}.  
Weak lensing peaks contain information on cosmological parameters including neutrino mass \citep{Ajani2020nuMassWLpeaks, DESY32022WLpeaks, Liu2023HSC_WLpeaks}.  
In addition, the spatial auto- and cross-correlations of galaxies, gravitational lensing and both thermal and kinetic SZ maps contain some degree of information about massive halos, and higher-order statistical signatures at non-linear scales are even more strongly connected.\footnote{For example, \href{https://www.snowmass21.org/docs/files/summaries/CF/SNOWMASS21-CF4_CF6_Jia_Liu-033.pdf}{this Snowmass2021 Letter of Interest}.}  
The spatial clustering of the cluster population itself is a signal that improves cosmological inference \citep{MajumdarMohr2004, Euclid2022clusterClustering}, and the power spectrum and bispectrum of massive halos contains potentially powerful information on primordial non-Gaussianities \citep{Coulton2023QuijotePNG}.   

Cluster counts offer complementary information to other cosmological probes, especially as the population is sensitive to both cosmic geometry and the gravitational growth of structure \citep{FriemanTurnerHuterer2008, Cunha2009}.  
A recent study that combines DES redMaPPeR cluster counts with spatial correlations of galaxy and lensing demonstrates the value of this approach \citep{To2021cluster2pt}. Clusters could be used to independently assess a recent CMB+LSS finding of a $4.2 \sigma$ larger than \lcdm\ growth factor index \citep{Nguyen2023}.  

These types of studies could potentially benefit from a compact mass function form, as DQ-HMF parameters could be used either as informative priors or as part of the focus of posterior likelihood evaluation.


\subsection{Other Caveats and Extensions}\label{sec:discExtensions}

We mention here a few additional caveats and potential extensions.  

\emph{Alternative Mass Conventions.}  In N-body simulations, the mass of a halo is typically defined by percolation or spherical overdensity approaches \citep{White2001haloMass}.  spherical  \citep[see, \eg][and references therein]{Diemer2020HMF}.  For the spherical overdensity approach, common choices for the interior mean density threshold and/or the reference density (critical or mean mass are typical choices) induce scale-dependent shifts in mass. The resultant HMF forms are follow similar forms, however, remain similar and can be converted using mean mass density profile shapes \citep[see Appendix B of][]{Evrard2002HubbleVolume}.  We suspect, but do not attempt to prove here, that a compact representation would be valid for most, if not all, existing conventions for true halo mass. 


\emph{Alternative Formulations for Extended Dynamic Range.}  Our model aims at near-field studies of groups and clusters.  To extend to model to lower-mass halos, one could include a transition mass scale below which the HMF would become pure power-law.  We note that the pure power law form at low masses ignores effects of baryon feedback during galaxy formation.  A recent internal structure study of halos across nearly six orders of magnitude in mass in the IllustrisTNG simulations \citep{Anbajagane2022DMscaling} finds wiggles in dark matter halo scaling relations near the Milky Way mass of $10^{12} \hinv \msol$, where star formation efficiency in the late universe peaks \citep{Behroozi2013SFH}.  This finding suggests that the HMF may also have a localized deviation from a pure power-law form at that scale.

The near-field halos above our chosen limiting mass of $10^{13.7} \hinv \msol$ comprise several percent of the overall matter density at $z<1.5$, but this fraction becomes negligible at much higher redshifts.  The mass scale associated with the most extreme few percent of the halo population declines with redshift, reaching Milky Way-scale halos that host bright galaxies at $z > 8$, as seen in JWST observations \citep{Boylan-Kolchin2023JWSTcounts}.  

To span a wider range in redshift, one could redesign the model by reframing the normalization.  Instead of the number density at fixed mass, $\beta_0(z)$, one could employ a mass scale at fixed number density parameter, for example, $M_{-6}(z)$ to represent the mass scale at which the comoving space density is $10^{-6}~h^3 \mpc^{-3}$.  To avoid cosmic volume uncertainties, the space density itself could be reframed in observable terms, in units of number per square degree per unit redshift.


\emph{Beyond binning.} The IM forecasts employ binned values for key sample characteristics of counts and mean mass. As multiple observable properties become available for larger population ensembles, a likelihood analysis that considers each system's true mass as additional model parameters \citep{Mulroy2019LoCuSS} could prove powerful.  

\emph{Multi-property statistics.}  The expressions derived in E14 for selection property-conditioned statistics still apply. We emphasize above only the mean mass and mass variance conditioned on the selection property, $s_a$, but expressions for one or more additional properties, $s_b$ (see equations (12) through (14) of E14) remain applicable, except now the HMF mass-shape parameters are explicitly redshift dependent, $\beta_i \rightarrow \beta_i(z)$. 


\section{Summary}\label{sec:summary} 

We introduce a compact representation for the differential space density of high mass halos that host groups and clusters and demonstrate its utility to match well the output of the Mira-Titan emulator of purely collisionless universes for masses $> 10^{13.7} \hinv \msol$ in the near cosmic field of redshifts $z<1.5$.  Convolving with a minimal MOR yields analytic forms for the space density and property-selected statistics that explicitly expose parameter degeneracies and that are fast to compute.  
Such a compact representation offers a common ground for cluster sample analysis independent of selection method.  

With roughly one million halos above $10^{14} \msol$ available on the full sky \citep{AllenEvrardMantz2011ARAA}, and studies of protoclusters at moderate redshifts in ascendancy \citep{AlbertsNoble2022protoclusters}, there is abundant information available from galaxy cluster surveys.  Unlocking that information will require careful modeling of sample selection, an endeavor that will be aided by sophisticated sky maps \citep[\eg][]{Schaye2023Flamingo}.  Near-term, more efforts to empirically study the MOR using high quality multi-wavelength data are needed. As the sample size of clusters with multiple well-measured properties grows from tens \citep[\eg][]{Mulroy2019LoCuSS} to hundreds \citep[\eg][]{Giles2022XCS+SDSSRM, Upsdell2023XCS+DES} to thousands, the detailed form of the multi-property MOR will come into focus, which can unlock more precise estimates of the underlying true mass of each system and, via collective study, the HMF and its behavior over cosmic time.  

\bigskip
\textsl{Acknowledgments.}  This work was initiated under NSF-REU Grant 2149884 and was also supported by NASA ADAP Grant 80NSSC-22K0476. This work employed open-source software of 
  {\sc NumPy} \citep{numpy}, {\sc SciPy} \citep{2020SciPy-NMeth}, and 
 {\sc Matplotlib} \citep{pyplot}.  
We dedicate this paper to the memory of Nick Kaiser, in honor of his seminal works on galaxy clusters, including spatial clustering \citep{Kaiser1984}, property scaling and evolution \citep{Kaiser1986SS, Kaiser1991}, gravitational lensing mass estimates \citep{KaiserSquires1993lensing, LuppinoKaiser1997MS1054lensing} and gravitational redshifts \citep{Kaiser2013MNRASgravRedshiftsClusters}. Part of this work was performed at the Aspen Center for Physics, which is supported by National Science Foundation grant PHY-1607611.

\bibliographystyle{mnras}
\bibliography{references.bib}


\appendix

\section{Three-parameter Toy Model}\label{sec:ToyModel}

We consider here a toy HMF model at a fixed redshift with only three degrees of freedom.  This model uses only two HMF and one MOR degree of freedom, so lessons learned here may not directly translate to the more complex, realistic cases presented in \S\ref{sec:Results}.  Nonetheless this simple example illustrates the value in having analytic forms for the information matrix. 

For this exercise, we consider a pure power-law mass function, with $\beta_2 \equiv 0$ and the normalization $\beta_0$ and slope $\beta_1$ the parameters of interest.  Along with these parameters, we consider the third degree of freedom to be the MOR variance, $\sigma^2$.  The parameter space with three degrees of freedom is thus $\vec{p} = \{ \beta_0, \beta_1, \sigma^2 \}$.  

To constrain these parameters we consider having three available measurements: i) a count of clusters above an observed property threshold; ii) an estimate of mean mass of these systems, and; iii) an estimate of the intrinsic mass scatter.  

For simplicity, we choose a linear MOR relation, $\alpha=1$, and use observed property units such that the normalization at the pivot mass scale is $\campi = 0$ (recall this is a log quantity).  We further simplify by choosing a threshold for counts at $s_{\rm min} = \campi = 0$.  We require that the HMF slope, $\beta_1$, be greater than one in order to have convergent counts.  

For this toy case, the counts above the minimum property threshold simplifies to 
\begin{eqnarray}\label{eq:countsSimple}
    N & = & \exp \left[ -\beta_0+ \frac{\beta_1^2 \sigma^2}{2} \right] \int_0^\infty ds\ \exp (-\beta_1 s) \\
   & = &  \beta_1^{-1}  \ \exp \left[ -\beta_0+ \frac{\beta_1^2 \sigma^2}{2} \right] .
\end{eqnarray}

The log-mean mass, equation~\eqref{eq:mubar}, reduces to $\langle \mu | s \rangle = s - \beta_1 \sigma^2$, and the mean mass at fixed $s$ is 
\begin{eqnarray}\label{eq:MsSimple}
 \langle M | s \rangle  & = & \int_{-\infty}^\infty d \mu \ \Pr (\mu | s) \ \exp(\mu)  \\
  & = & \int_{-\infty}^\infty d \mu \ \exp \left[ - \frac{ \left( \mu - (s- \beta_1 \sigma^2) \right)^2}{2 \sigma^2} + \mu  \right] \\   
   & = &  \exp \left[ s - \left(\beta_1 - \frac{1}{2} \right) \sigma^2 \right]  .
\end{eqnarray}
The mean mass of the thresholded population is the number-weighted value
\begin{eqnarray}\label{eq:MSimple}
 \langle M \rangle & = & 
 \frac{ \int_0^\infty ds \ \langle M | s \rangle \ n(s) }{ \int_0^\infty ds \ n(s) } \\
  & = & \exp\left[ - \left( \beta_1 - \frac{1}{2} \right) \sigma^2 \right] \frac{ \int_0^\infty ds \exp \left( -(\beta_1-1) s \right)}{\int ds_0^\infty \exp \left( -\beta_1 s \right)} \\
   & = &  \frac{\beta_1}{\beta_1-1} \exp\left[ -\left(\beta_1 - \frac{1}{2}\right) \sigma^2 \right]  .
\end{eqnarray}
We assume a fractional uncertainty, $\errmubar$, in this estimate, meaning we assign this uncertainty to the logarithm, 
\begin{equation}\label{eq:LMMsimple}
 \ln  \langle M \rangle =  - \left(\beta_1 - \frac{1}{2}\right) \sigma^2 \ + \ \ln \left(  \frac{\beta_1}{\beta_1-1}  \right) .
\end{equation}

When $\alpha = 1$ and $\beta_2=0$, the mass variance at fixed property, equation~\eqref{eq:muVar}, is identical to the fractional property variance at fixed mass.  Since we again assign a fractional uncertainty, $\errVarmu$, to the variance the relevant expression is again logarithmic,   
\begin{equation}\label{eq:VARsimple}
\ln \sigmamu^2 \ = \ \ln \sigma^2 . 
\end{equation}

\subsection{IM analysis}\label{sec:ToyModelIM}

The three pieces of information relevant for the IM analysis are: i) the count, $N$, equation~\eqref{eq:countsSimple}; ii) a fractional error, $\errmubar$, on the mean mass, equation~\eqref{eq:LMMsimple}, and; iii) a fractional error, $\errVarmu$, on the mass variance, equation~\eqref{eq:VARsimple}.  

The information matrix is the sum of three contributions 
\begin{equation}\label{eq:Fsimple}
\mathcal{F}  = \mathcal{F}_C \ + \ \mathcal{F}_{M} \ + \ \mathcal{F}_{V} ,
\end{equation} 
from counts, mean mass, and mass variance, respectively. For clarity we write only the upper half of the symmetric matrix with rows and columns in the order $\{ \beta_0, \beta_1, \sigma^2 \}$.  

The first term represents the contribution from the expected counts under Poisson statistics, 
\begin{equation}\label{eq:FsimpleCount}
\mathcal{F}_C = 
\begin{bmatrix}
1 & \beta_1^{-1} -\beta_1 \sigma^2 & -\beta_1^2/2   \\[4truept]
 & (\beta_1 \sigma^2 -\beta_1^{-1})^2 &  (\beta_1^3 \sigma^2 - \beta_1)/2 \\[4truept]
 & & \beta_1^4/4
\end{bmatrix}
\ N .
\end{equation}

The second term represents the contribution from measuring the mean mass, 
\begin{multline}\label{eq:FsimpleMbar}
\mathcal{F}_M = \ \frac{1}{\errmubar^2} \\
\begin{bmatrix}
0 & 0 & 0  \\[4truept]
 & \left( \sigma^2 + [\beta_1^2-\beta_1]^{-1} \right)^2  & (\beta_1-\frac{1}{2}) \left( \sigma^2 + [\beta_1^2-\beta_1]^{-1} \right) \\[4truept]
 & & (\beta_1- \frac{1}{2}) ^2
\end{bmatrix}
 .
\end{multline}

The final term represents the contribution from the mass variance 
\begin{equation}\label{eq:FsimpleMVar}
\mathcal{F}_V = 
\begin{bmatrix}
0 & 0 & 0  \\[4truept]
 & 0 & 0 \\[4truept]
 & & \frac{1}{\sigma^4} 
\end{bmatrix}
\ \frac{1}{\errVarmu^2} .
\end{equation}

For this toy example, we imagine a low redshift sample covering sufficient sky area with sensitivity sufficient to acquire a property-limited sample of 2500 clusters.  
For this sample size, the Poisson limiting error in the HMF amplitude, $\beta_0$ is $0.02$.  This value is reached in the limit of zero uncertainties in the other measurements, $\errmubar \rightarrow 0$ and $\errVarmu \rightarrow 0$.

\begin{table}\label{tab:simplecases}
	\centering
\begin{tabular}{ c | c  c | c  c } 
  & \multicolumn{2}{c|}{Input Values} &  \multicolumn{2}{c}{Output Constraints}   \\ 
 Case & \errmubar & \errVarmu & $\varepsilon (\beta_0$) & $\varepsilon (\beta_1$)  \\ 
 \hline
 I & 0.1 & 0.5 & 0.15 & 0.21 \\ 
 II & 0.01 & 0.5 & 0.14 & 0.13 \\
 III & 0.1 & 0.05 & 0.056 & 0.17 \\ 
 IV & 0.01 & 0.05 & 0.025 & 0.021 \\ 
\end{tabular}
\caption{IM results on the normalization and slope in four $3 \times 3$ cases.  The second and third columns list the errors on mean mass and mass variance assumed for each case, while the last two columns list forecast errors on the normalization, $\beta_0$, and slope, $\beta_1$.  All models use total counts of $N=2500$, implying a Poisson-limited constraint of $0.02$ on the HMF amplitude, $\beta_0$.  The constraint on $\sigma^2$ is not improved by the information available, see equation~\eqref{eq:FsimpleSigma2}.
} 
\end{table}

The determinant of the information matrix, which measures the information volume, has the form
\begin{equation}\label{eq:FsimpleDet}
\det (\mathcal{F}) = \left( 1 + \frac{1}{\sigma^2 \beta_1(\beta_1-1)} \right)^2 N \errmubar^{-2} \ \errVarmu^{-2}  .
\end{equation} 
The information volume increases with larger counts or smaller fractional systematic uncertainties. We assume a slope value, $\beta_1=2$, appropriate for \lcdm\ at $z \simeq 0.2$, along with a property variance at fixed mass, $\sigma^2 = 0.1 \simeq 0.3^2$.   Note that the HMF slope is sufficiently steep to avoid the singular case, $\beta_1 = 1$.  For this set of parameters the square prefactor of equation~\eqref{eq:FsimpleDet} takes on a value of 36.  

Inverting the information matrix yields the anticipated parameter constraints.  The 3-3 element, providing the expected error on the property variance, $\varepsilon (\sigma^2)$, is simply
\begin{equation}\label{eq:FsimpleSigma2}
\varepsilon (\sigma^2) \equiv  \left( \mathcal{F}^{\, -1} \right)_{33}^{1/2} = \errVarmu \sigma^2  .
\end{equation} 
This result follows from the fact that our choices of $\alpha = 1$ and $\beta_2=0$ imply that the mass variance and MOR variance are equal (see equation~\eqref{eq:VARsimple}).   This result also means that information from the counts and mean mass is decoupled from the property variance dimension, despite the fact that $\sigma^2$ is involved in the IM contributions from both the counts and mean mass.  


We consider two levels of fractional uncertainty for each of the mean sample mass and the mass variance, leading to the four cases shown in Table~\ref{tab:simplecases}.  Baseline uncertainties (Case I) are $0.1$ and $0.5$, respectively, while the optimistic case (IV) improve on these values by an order of magnitude.  Cases II and III separately use optimistic values for the mean mass and mass variance, respectively, holding the other quality parameter at the baseline level.  

The forecast uncertainties in the model parameters for each case are listed in Table~\ref{tab:simplecases}.  For case I, the baseline, an uncertainty of $0.15$ is expected on the HMF amplitude, $\beta_0$ --- nearly eight times the Poisson limit --- with a similar constraint of $0.21$ on the HMF slope, $\beta_1$.  

For Case II, in which the uncertainty in log-mean mass is reduced by an order of magnitude, the extra constraining power only modestly improves the slope uncertainty, to $0.13$.  The normalization uncertainty remains nearly unchanged from Case I.  The log-mean mass, equation~\eqref{eq:LMMsimple}, involves the product, $\beta_1 \sigma^2$, so the weak constraint on $\sigma^2$ limits how well $\beta_1$ can be recovered even with a 1\% measurement.  In turn, this weakness propagates into a poor constraint on the normalization, $\beta_0$, via the counts, equation~\eqref{eq:countsSimple}.

Case III demonstrates the utility of better understanding of the mass variance.  When the error in $\sigmamu^2$ is reduced by an order of magnitude, a substantially better result, with uncertainty $0.056$, is obtained for the amplitude, $\beta_0$, a reduction of nearly a factor of three relative to Case I.  The slope uncertainty of $0.17$ lies intermediate between that of Cases I and II.  

Finally, Case IV improves both the log-mean mass and mass variance uncertainties.  The result is a nearly Poisson-limited constraint on the HMF amplitude, $\varepsilon(\beta_0) = 0.025$, along with a slightly smaller uncertainty, $0.021$, on the slope.   

From this exercise, one might infer that the HMF normalization is more sensitive to knowledge of mass variance than knowledge of mean mass.  But the simplicity of our chosen case may be misleading, as real survey applications are more complex,
involving more terms of counts and mean mass than mass variance.
Survey-specific analysis must be performed to understand the relative benefits of these sources of systematic error.  However, the Gold Sample forecast of \S\ref{sec:discSelectionML} illustrates the utility of a lower variance mass proxy over counts in a practical survey application.   




\end{document}